\documentclass[%
 reprint,
 amsmath,amssymb,
 aps,
]{revtex4-1}

\usepackage{graphicx}
\usepackage{multirow}
\usepackage{dcolumn}
\usepackage{bm}
\usepackage[colorlinks=true,linkcolor = blue,citecolor = red,urlcolor  = blue,anchorcolor = blue]{hyperref}

\usepackage[english]{babel}

\begin{document}
\def\red#1{{\color{red}#1}}

\title{Quantum transport senses community structure in networks}

\author{Chenchao Zhao}
\author{Jun S.~Song}%
 \email{songj@illinois.edu}
\affiliation{%
Department of Physics, University of Illinois at Urbana-Champaign,
Urbana, IL
}
\affiliation{%
Carl R. Woese Institute for Genomic Biology, University of Illinois at Urbana-Champaign,
Urbana, IL
}
%


\begin{abstract}

  Quantum time evolution exhibits rich physics, attributable to the
  interplay between the density and phase of a wave function.
  However, unlike classical heat diffusion, the wave nature of quantum
  mechanics has not yet been extensively explored in modern data
  analysis.  We propose that the Laplace transform of quantum
  transport (QT) can be used to construct an ensemble of maps from a
  given complex network to a circle $S^1$, such that closely-related
  nodes on the network are grouped into sharply concentrated clusters
  on $S^1$.  The resulting QT clustering (QTC) algorithm is as
    powerful as the state-of-the-art spectral clustering in discerning
    complex geometric patterns and more robust when  clusters show strong density
    variations or heterogeneity in size.  The observed phenomenon of
  QTC can be interpreted as a collective behavior of the microscopic
  nodes that evolve as macroscopic cluster ``orbitals'' in an
  effective tight-binding model recapitulating the network. 
  Python source code implementing the algorithm and
  examples are available at \url{https://github.com/jssong-lab/QTC}.

\end{abstract}

\pacs{Valid PACS appear here}
\maketitle


Grouping similar objects into sets is a fundamental task in modern
data science.  Many clustering algorithms have thus been devised to
automate the partitioning of samples into clusters, or communities,
based on some similarity or dissimilarity measures between the samples
that form nodes on a graph \cite{Kaufman:2009dk,Hastie:2013fd}.  In
particular, physics-inspired approaches based on classical spin-spin
interaction models \cite{Li:2012ks,Reichardt:2004ea} and Schr\"odinger
equation \cite{Horn:2001dc} have been previously proposed; however,
the former usually requires computationally intensive Monte Carlo
simulations which may get trapped in local optima, while the latter
essentially amounts to Gaussian kernel density estimation.  These
intriguing physical ideas thus have been under the shadow of popular
contemporary approaches that are simple and computationally efficient,
such as the dissimilarity-based KMeans
\cite{Lloyd:1982do,Forgy:1965eu,Arthur:2007tv} and hierarchical
clustering \cite{Sibson:1973jl,Defays:1977ix}, density-based DBSCAN
\cite{Ester:1996tm}, distribution-based Gaussian mixture
\cite{Yu:2012ew}, and kernel-based spectral clustering
\cite{vonLuxburg:2007bb}.  By contrast, we here use the physics of
quantum transport (QT) on data similarity networks to devise a simple
and efficient algorithm. The performance of QT clustering (QTC) is
comparable to the state-of-the-art spectral clustering when the
clusters exhibit non-spherical, geometrically complex shapes; at the
same time, QTC is less sensitive to the choice of parameters in the
kernel.  Moreover, unlike spectral clustering, the QT representation
of data on a circle does not jump in dimension when the specified
number of clusters changes.

Heat diffusion has been applied to rank web page popularity
\cite{Brin:1998jv}, probe geometric features of data distribution
\cite{Coifman:2006cy}, and measure similarity in classification
problems \cite{Lafferty:2015uy,Zhao:2017wla}.  By contrast, despite
the formal resemblance between the heat equation and the Schr\"odinger
equation, the time evolution of a quantum wave function has been
largely ignored in machine learning.  Both heat and Schr\"odinger
equations have conserved currents; however, while the heat current is
proportional to the negative gradient of heat density itself, the
velocity of quantum probability current is set by the phase gradient
which satisfies the Navier-Stokes equation, making quantum probability
density an irrotational fluid (Supplemental Material (SM)
\footnote{see online Supplemental Material for detailed derivations
  and additional information.}, I). Thus, the Schr\"odinger equation
embodies richer physics than heat diffusion and can capture
spatiotemporal oscillations and wave interference. One promising
observation has been that quantum time evolution can be faster in
reaching faraway nodes compared with heat diffusion in ordered binary
tree networks, suggesting the possibility of finding practical
applications of quantum mechanics in network analysis
\cite{Farhi:1998ez,Childs:2002kj,Shenvi:2003be,Faccin:2013gq}.
However, there are several outstanding challenges: e.g., unlike the
heat kernel, the oscillatory quantum probability density is monotonic
in neither time nor spatial distance; moreover, 
irregularities in either edge weights or network structure can
severely restrict the propagation of a wave function through
destructive interference, analogous to Anderson localization in
disordered media \cite{Anderson:1958fza}.  We circumvent these
difficulties associated with using the probability density itself and
demonstrate the utility of the phase information for clustering network nodes.

A generic undirected weighted network, e.g.~a data similarity network
of $m$ samples in $\mathbb{R}^d$ represented as nodes, is encoded by
an $m\times m$ symmetric adjacency matrix $A$. The row or column sum
vector ${\rm deg}(i) = \sum_k {A_{ik}} = \sum_k {A_{ki}}$ gives rise
to the diagonal degree matrix $D={\rm diag}({\rm deg})$.  Replacing
the continuous Laplacian with the graph Laplacian $L=D-A$ then
discretizes the heat and Schr\"odinger equations on data similarity
networks. Enforcing the conservation of discrete heat current
introduces the normalized graph Laplacian $Q = L D^{-1}$.  The
original graph Laplacian $L$ of an undirected network is automatically
Hermitian, but we adopt the symmetrized version
$H=D^{-\frac12}LD^{-\frac12}$ of $Q$ as our Hamiltonian, since it has
the same spectrum as $Q$. With this choice, $H$ has a nontrivial
ground state $\psi_0 (i) \propto \sqrt{{\rm deg}(i)}$
\cite{Faccin:2013gq}.

\begin{figure}[thb]
    \centering
    \includegraphics[width=1.0\columnwidth]{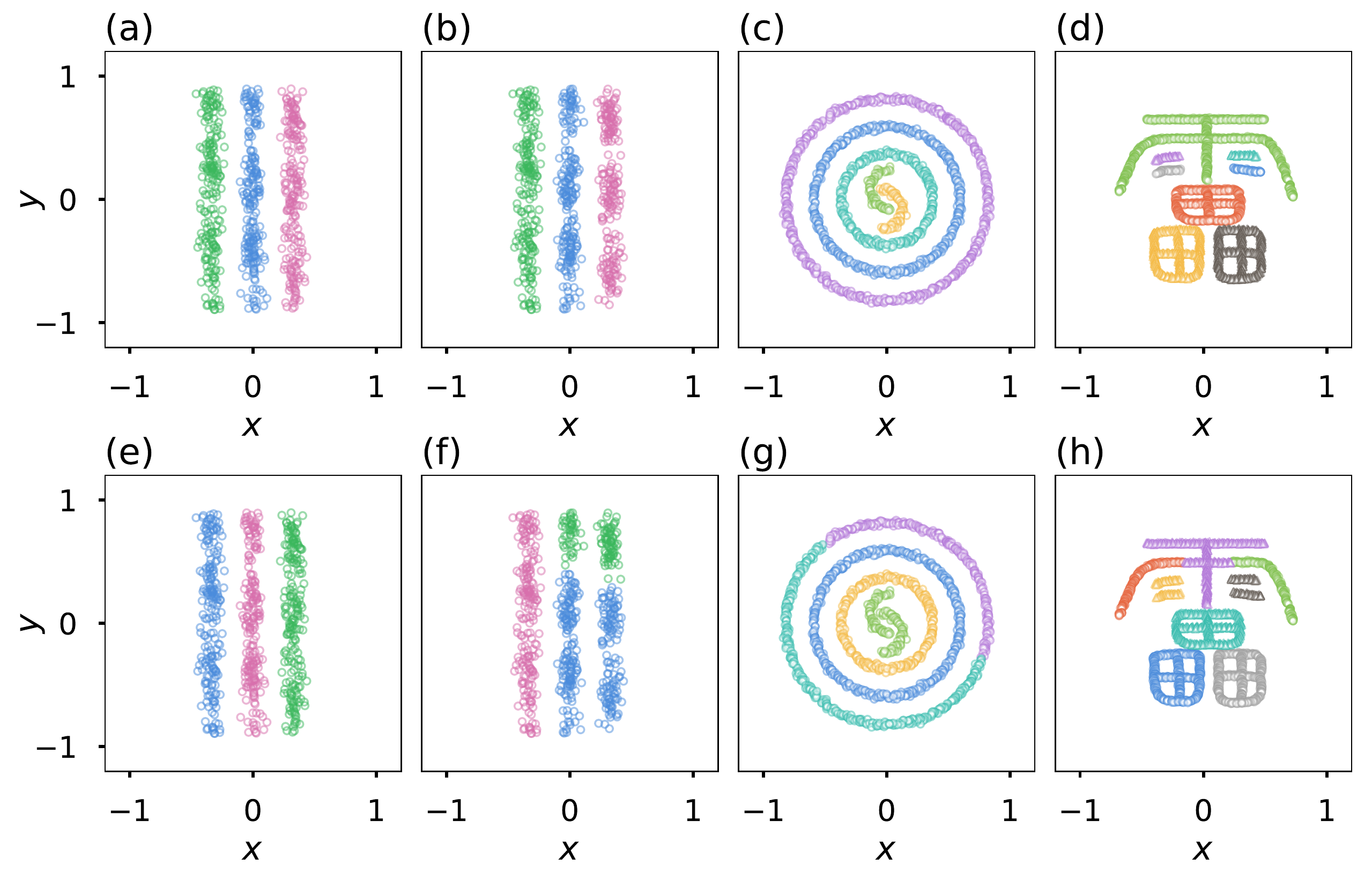}
    \caption{Comparison of (a-d) QTC and (e-h) spectral clustering
      using synthetic data. We specified three clusters for (a,b,e,f),
      five clusters for (c,g), and eight clusters of (d,h).  We
        chose intermediate values of proximity measure $r_\varepsilon$ in the Gaussian
        similarity function to demonstrate the robustness of
        QTC; spectral clustering was able to produce the correct clustering
        only when $r_\varepsilon$ was tuned to be
        sufficiently small.}
    \label{fig:syn_examples}
\end{figure}

\begin{figure}[!thb]
    \centering
    \includegraphics[width=1.0\columnwidth]{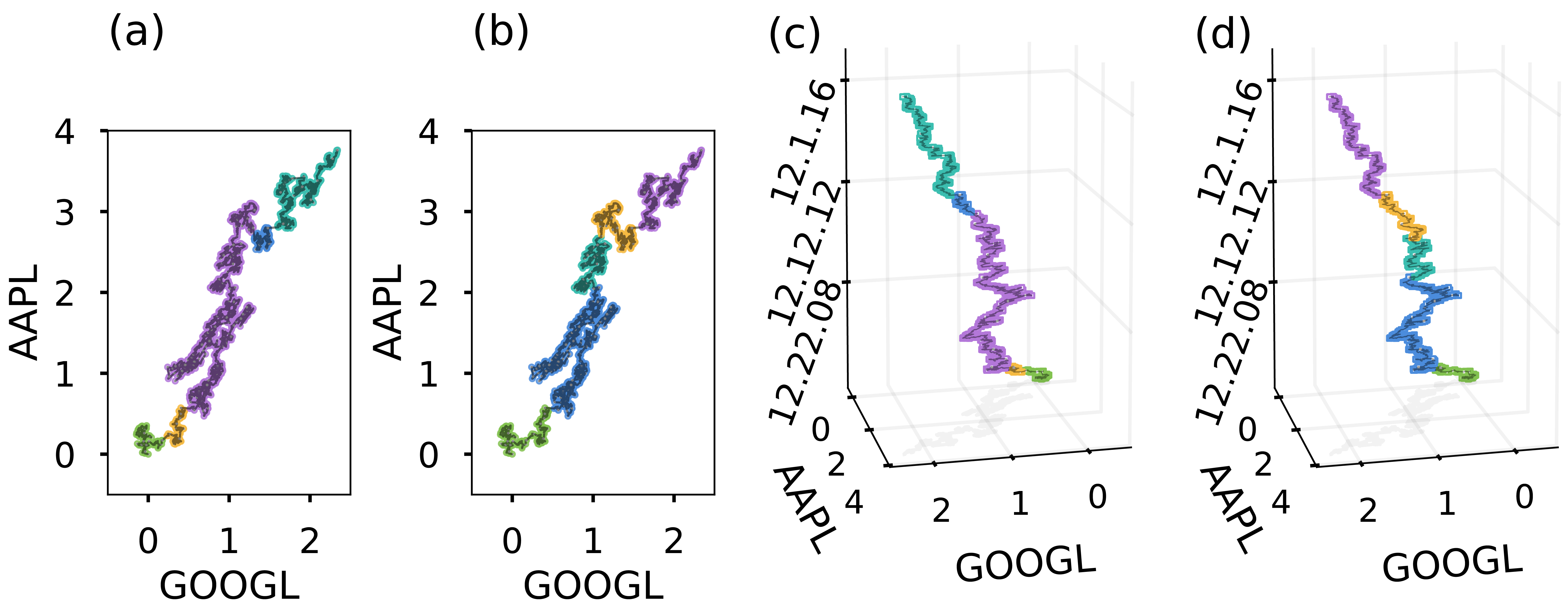}
    \caption{Comparison of (a) QTC and (b) spectral clustering using the time series data of
      log-prices of \textsc{aapl} and \textsc{googl} stocks from January 3, 2005 to
      November 7, 2017.  Five clusters were specified, and the $1\%$-quantile $r_{1\%}$ was chosen as the proximity measure.  The time
      evolution trajectories of data in (a) and (b) are displayed in
      (c) and (d), respectively, with an extra temporal dimension.  }
    \label{fig:timeseries}
\end{figure}

\begin{table}[htb]
\begin{centering}
\caption{Daily returns (\%) at the identified jumps in Fig.~\ref{fig:timeseries} }
\label{tab:timeseries_jumps}
\begin{tabular}{c|c|c|c|c|c|c|c|c}
\hline\hline 
\multirow{2}{*}{Date} & \multicolumn{2}{c|}{2005} & \multicolumn{3}{c|}{2010} & 2012 & \multicolumn{2}{c}{2013}\tabularnewline
\cline{2-9} 
& 5/23 & 10/21 & 4/16 & 4/20 & 4/21 & 2/8 & 1/24 & 10/18\tabularnewline
\hline 
 \textsc{googl} & $+5.6$ & $+11.4$ & $-7.9$ & $+0.9$ & $-0.1$ & $+0.5$ & $+1.7$ & $+13.0$\tabularnewline
\hline 
\textsc{aapl} & $+5.7$ & $-0.8$ & $-0.6$ & $-0.1$ & $+5.8$ & $+1.7$ & $-13.2$ & $+0.9$\tabularnewline
\hline 
\multicolumn{1}{c}{} & \multicolumn{1}{c}{Q} & \multicolumn{1}{c}{Q S} & \multicolumn{1}{c}{S} & \multicolumn{1}{c}{S} & \multicolumn{1}{c}{S} & \multicolumn{1}{c}{S} & \multicolumn{1}{c}{Q} & Q S\tabularnewline
\hline \hline
\end{tabular}
\end{centering}
\end{table}

For concreteness, we define the pairwise
similarity or adjacency between sample ${\bf x}_i$ and sample ${\bf
  x}_j$ by the Gaussian function $A_{ij} =
\exp(-r_{ij}^2/r_\varepsilon^2)$, where $r_{ij}=\lVert{\bf x}_i - {\bf
  x}_j \rVert$ is the Euclidean distance and $r_\varepsilon$ is the
$\varepsilon$-quantile among $r_{ij}>0$.  Ideally, the proximity
measure $r_\varepsilon$ is chosen such that for samples $i$ and $j$
belonging to distinct clusters, we have $r_{ij} \gg r_\varepsilon$,
but within any given cluster, a pair $(i,j)$ of nearest neighbors has
$r_{ij} \sim {\cal O}(r_\varepsilon)$.

Defining the Laplace transform of a wave function initially localized
at node $j$ and evaluated at node $i$ as \cite{Note1,Anderson:1958fza}
\begin{equation}
  {\cal L}[\psi(i|j)](s) = \int_0^\infty dt\, \langle i| {\rm e}^{-{\rm i}Ht -st}|j\rangle,
\end{equation}
our clustering algorithm stems from the observation that the phase
$\Theta(i|j)$ of this transformed function is essentially constant as
$i$ varies within a cluster, but jumps as $i$ crosses clusters (see
discussion below; \cite{Note1}). The phase information thus provides a
one-dimensional representation of data on $S^1$, such that distinct
clusters populate separable regions on $S^1$; intuitively, the phase
distribution $\Theta(\cdot|j)$ corresponds to a specific perspective
on community structure sensed by the wave packet initialized at node
$j$.  In general, the phase distribution $\Theta(\cdot|j)$ changes
with the initialization node $j$.  Thus, if we randomly choose $m'$
initialization nodes ($m'\approx 100$ for data sets in Fig.~1 \& 2),
for $1< m'\le m$, then we obtain an ensemble of $m'$ phase
distributions, in each of which the phase is almost constant within
clusters; this ensemble ultimately provides a collection of
perspectives on the underlying community structure, as sensed by the
wave packets initialized at the chosen nodes.

In practice, we {\em a priori} specify the number $q$ of clusters, and
use the phase distribution of each wave function to partition the
nodes into $q$ subsets \cite{Note1}.  We label each of the $m''$
distinct partitions by an integer $\alpha$, where $m''\le m'$, and
calculate the occurrence frequency $w_\alpha \in (0,1]$ of each
partition, such that the normalization condition $\sum_{\alpha}
w_\alpha = 1$ holds (SM \cite{Note1}, I C).  Typically, we find that
the frequencies are dominated by a single partition; other $m''-1$
less frequent partitions may arise from wave functions initialized at
nodes of a small subnetwork isolated from the rest of the network.
Hence, the minority predictions provide less holistic views of the
network community structure, and we choose the majority prediction
from the ensemble as our final clustering decision.

We compared the performance of QTC to spectral clustering
\footnote{When without an explicit specification, the affinity matrix
  used in spectral clustering is the same one used in QTC.} using four
synthetic data sets having complex geometry
(Fig.~\ref{fig:syn_examples}): (1) uniform sticks, (2) non-uniform
sticks, (3) concentric annuli, and (4) the Chinese character for
``thunder.''  Both algorithms performed equally well on the simple
data set of uniformly sampled sticks
(Fig.~\ref{fig:syn_examples}(a,e)) or when $r_\varepsilon$ was chosen
to be sufficiently small such that the clusters became almost disjoint
subnetworks; as $r_\varepsilon$ increased, however, QTC remained
robust (Fig.~\ref{fig:syn_examples}(b-d)), while spectral clustering
made mistakes (Fig.~\ref{fig:syn_examples}(f-h)).
We further tested QTC on time-series stock price data (data
preparation methods in SM \cite{Note1}, II).  The log-prices of a
portfolio of stocks form a random walk in time with occasional jumps
which are often triggered by important events such as the release of
fiscal reports and sales records.  The jumps then separate the
fractal-like trajectory of historical log-prices into several
performance segments.  Figure \ref{fig:timeseries}(a,b) shows the
log-price distribution of two stocks, \textsc{aapl} and
\textsc{googl}, from January 3, 2005 to November 7, 2017, where we removed
the temporal information from the data set. 
When we specified five clusters,
QTC cut the trajectory into five consecutive segments in the temporal
space (Fig.~\ref{fig:timeseries}(a,c)) with heterogeneous lengths, whereas spectral clustering
partitioned the trajectory into clusters of similar sizes and mixed the temporal
ordering near the boundary of blue and cyan clusters
(Fig.~\ref{fig:timeseries}(b,d)).  The jumps identified by QTC (Q's in
Table \ref{tab:timeseries_jumps}) coincided with major news events for
the two stocks, whereas spectral clustering (S's in Table
\ref{tab:timeseries_jumps}) failed to identify the large drop of
\textsc{aapl} on 1/24/2013 and instead included several less
significant stock movements.
These results showed that QTC was more robust than the conventional
spectral embedding method on non-spherical data distributions with
anisotropic density fluctuations (Fig.~\ref{fig:syn_examples}(b,f)) or
complex geometric patterns exhibiting a hierarchy of cluster sizes
(Fig.~\ref{fig:syn_examples}(c,g) and (d,h);
Fig.~\ref{fig:timeseries}).

Next, we provide a physical interpretation of the agglomeration
phenomena observed in QTC using an effective tight-binding model. For this
purpose, we rewrite the Laplace transform as ${\cal L}[\psi(i|j)](s)
\equiv {\rm i}G (i,j; {\rm i}s)$, where
\begin{equation}
G(i,j; z) \equiv  \langle i | (z-H)^{-1} |j\rangle  
	= \sum_{n=0}^{m-1} \frac{\langle i | \psi_n\rangle \langle \psi_n | j \rangle}{z - E_n}\label{eq:resolvent_sum}	
\end{equation}
is the resolvent of $H$, and $\psi_n$ and $ E_n$ are the eigenvectors
and eigenvalues of $H$, respectively, for $n=0,1,\ldots,m-1$. We
assume that $E_n$ are ordered in a non-decreasing way.  As a result of
our choice of short-proximity adjacency measure, the largest
contributions to ${\rm i}G(i,j;{\rm i}s)$ come from the low energy
collective modes in the case of well-separated $q$ clusters indexed by
$\mu=0,1,\ldots, q-1$.  In this case, the ground state density
$|\psi_0(i)|^2 \propto {\rm deg}(i)$ will be accumulated around the
hub nodes within each cluster.  Furthermore, $H$ is essentially
$q$-block diagonal upon relabeling the nodes and exhibits a large
energy gap separating the low energy collective modes
$\{|\psi_n\rangle\}_{0\le n<q}$ from the high energy eigenstates
$\{|\psi_n\rangle\}_{q\le n < m}$ capturing microscopic fluctuations
within each cluster.  Notice that the major contribution to the
  resolvent in Eq.~\ref{eq:resolvent_sum} comes from terms with $n<q$,
  and that the number of low energy states equals the number of
  well-separated clusters (SM \cite{Note1}, I B
  and Fig.~S1). These observations thus motivate
  a $q$-dimensional coarse-grained Hamiltonian describing only
  the low energy collective modes.

Let $\{\phi_\mu\}_{\mu=0}^{q-1}$ be the cluster wave functions, or
``atomic orbitals,'' satisfying $\phi_\mu(i) > 0$ for $i$ in cluster
$\mu$ and zero elsewhere, and $\langle \phi_\mu | \phi_\nu \rangle =
\delta_{\mu\nu}$.  The effective tight-binding Hamiltonian is
\begin{equation}
	\hat H \equiv \sum_{\mu,\nu = 0}^{q-1} h_{\mu\nu} |\phi_\mu \rangle \langle \phi_\nu |,\; \text{and}\; h_{\mu\nu} \equiv  \xi_\mu \delta_{\mu \nu} +  v_{\mu\nu}, 
\end{equation}
where $\xi_\mu= \langle \phi_\mu | H | \phi_\mu \rangle$ describes the
ground state energy of each $\phi_\mu$, and the off-diagonal matrix
$v_{\mu\nu}= \langle \phi_\mu | H | \phi_\nu \rangle \mbox{ for
}\mu\neq\nu$, with $v_{\mu\mu}=0$, couples the atomic orbitals $\phi_\mu$ and $\phi_\nu$.
Through the diagonalization of the tight-binding Hamiltonian
$h_{\mu\nu}$, the $q$ atomic orbitals are then linearly combined into
$q$ molecular orbitals.

To illustrate the effects of off-diagonal coupling, we split $\hat H$
into diagonal $\hat H_0$ and off-diagonal $\hat V$, and study the Born
approximation of the Lippmann-Schwinger equation
\begin{equation}
\hat G(z) = \hat G_0(z) + \hat G_0(z) \hat V \hat G(z),
\end{equation}
where $\hat G(z) = (z-\hat H)^{-1}$ and $\hat G_0(z) = (z-\hat H_0)^{-1}$.
The effective resolvent matrix can thus be expanded as
\begin{equation}
g_{\mu\nu}(z)
=\frac{\delta_{\mu\nu}}{z-\xi_\mu} +
\frac{v_{\mu\nu}}{(z-\xi_\mu)(z-\xi_\nu)} + {\cal O}(v^2),
 \label{eq:Born}
\end{equation}
which is a weighed sum over all tunneling paths from cluster $\mu$ to
$\nu$, and converges quickly if $|v_{\alpha\beta}|\ll |z-\xi_\beta|$
for all $\alpha,\beta = 0, 1, \ldots, q-1$ (SM \cite{Note1}, I D,
Eq.~S3).  The propagator from node $j$ to $i$ in the effective
tight-binding theory, approximating Eq. \ref{eq:resolvent_sum}, is
directly related to $g_{\mu\nu}(z)$ as
\begin{equation}
g(i,j;z) = \sum_{\mu,\nu=0}^{q-1} \phi_{\mu}(i) g_{\mu\nu}(z)\phi^*_{\nu}(j).
\end{equation}
If the nodes $i$ and
$j$ belong to two  non-overlapping clusters $\mu$ and $\nu$,
respectively, then the propagator reduces to $g(i,j;z)=\phi_\mu(i)\phi_\nu(j)
g_{\mu\nu}(z)$ and $\arg g(i,j;z) = \arg g_{\mu\nu}(z)$, because of the
disjoint support and the non-negativity of cluster wave
functions.
In other words, the propagator initiated at $j$ has
a constant phase at all nodes $i$ within each cluster, and the phase associated
with each cluster is completely determined by the phase of resolvent matrix
$g_{\mu\nu}$, which in turn depends on the weak coupling $v_{\mu\nu}$
via Eq.~\ref{eq:Born}.

\begin{figure}
    \centering
    \includegraphics[width=1.0\columnwidth]{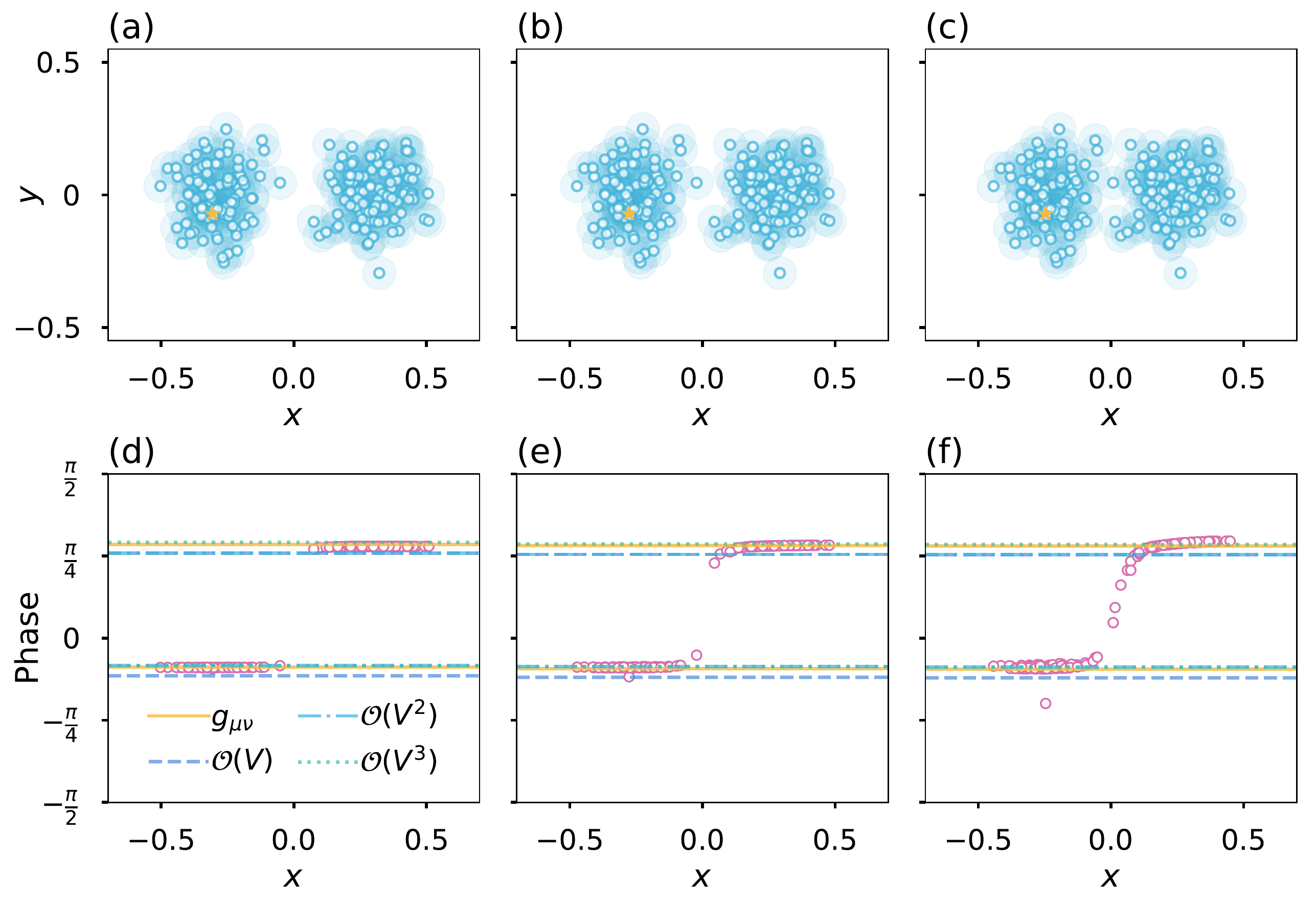}
    \caption{Two Gaussian clouds from ${\cal N}((\pm \ell, 0)^\top,
      \sigma^2{\bf 1}_{2\times 2} )$ with variations in the
      center-to-center distance (a) $\ell=3\sigma,$ (b)
      $\ell=2.7\sigma,$ and (c) $\ell=2.4\sigma$. Adjacency matrices
      were calculated using $r_\varepsilon = \sigma$. The radius of
      the faint large circle around each data point indicates
      $r_\varepsilon/2$. (d-f) The phase distributions (red circles) of all sample
      points  from (a-c), respectively; exact theoretical
      predictions $\arg \{{\rm i} g_{\mu\nu}({\rm i}s) \}$ from the
      low-energy effective model (solid line); the first, second, and
      third order perturbative approximations (dashed lines).  The
      Laplace transform parameter was set to $s=1.2(E_1 - E_0)$.  The
      $\star$ in (a), (b), and (c) mark the initialization nodes.  }
    \label{fig:two_clouds}
\end{figure}

As an example, consider two sets of $m$ samples drawn from ${\cal
  N}((\pm \ell, 0)^\top, \sigma^2{\bf 1}_{2\times 2} )$, respectively.
The effective 2-level Hamiltonian and resolvent matrices are
\begin{equation}
    h = \begin{pmatrix}
    \xi_0 & v \\
    v & \xi_1
    \end{pmatrix},
\ {\rm and }\
g(z) = \begin{pmatrix}
    z-\xi_0 & - v \\
    - v & z-\xi_1
    \end{pmatrix}^{-1}.
\end{equation}
As we vary $\ell=3\sigma, 2.7\sigma,$ and $2.4\sigma$, with a fixed
proximity length scale $r_\varepsilon = \sigma$, the cluster
configuration ranges from (a) well-separated, (b) in proximity, and
(c) overlapping (Fig.~\ref{fig:two_clouds}; Fig.~S2).  For each case,
Fig.~\ref{fig:two_clouds}(d-f) show the phase distribution of all
samples when quantum transport is initialized at one of the nodes in
the left cluster; it is seen that our theoretical prediction $\arg
\{{\rm i} g_{\mu\nu}({\rm i}s) \}$ and its perturbative approximations
calculated from Eq.~\ref{eq:Born} agree well.  Furthermore, if the two
clusters are identical, i.e.~$\xi_0=\xi_1$, then the effective 2-level
model can be mapped to the classic double-well tunneling model (SM
\cite{Note1}, I E); in this case, the phase distribution of the
Laplace transform of exact instanton solution matches that of our
simulated Gaussian clouds (Fig.~S3(a)).  When the weak coupling
assumption is not satisfied, the low-energy theoretical predictions
serve only as asymptotic limits, and some ambiguous points in a
strongly mixed region may have a phase that interpolates between the
theoretical predictions (Fig.~\ref{fig:two_clouds}(c,f); Fig.~S4 \&
S5).

When the clusters in data show strong mixing, no single partition may
be clearly dominant, so using the partition corresponding to the
highest occurrence frequency $w_\alpha$ may be unstable. In this
scenario, we propose a ``fuzzy'' summary of the ensemble.  Across $m'$
different initializations, we count the number of times where two
nodes, say $i$ and $k$, are assigned to the same cluster, and then
divide the count by $m'$.  We thereby arrive at a symmetric consensus
matrix $C_{ik}$ with 1 along the diagonal and other entries in $[0,1]$
(SM \cite{Note1}, I C).  The consensus matrix
provides a useful visualization of processed clustering structure and
also serves as a new input similarity measure suitable for many
popular statistical learning algorithms, such as spectral clustering,
hierarchical clustering, and SVM.

\begin{figure}
    \centering
    \includegraphics[width=1.0\columnwidth]{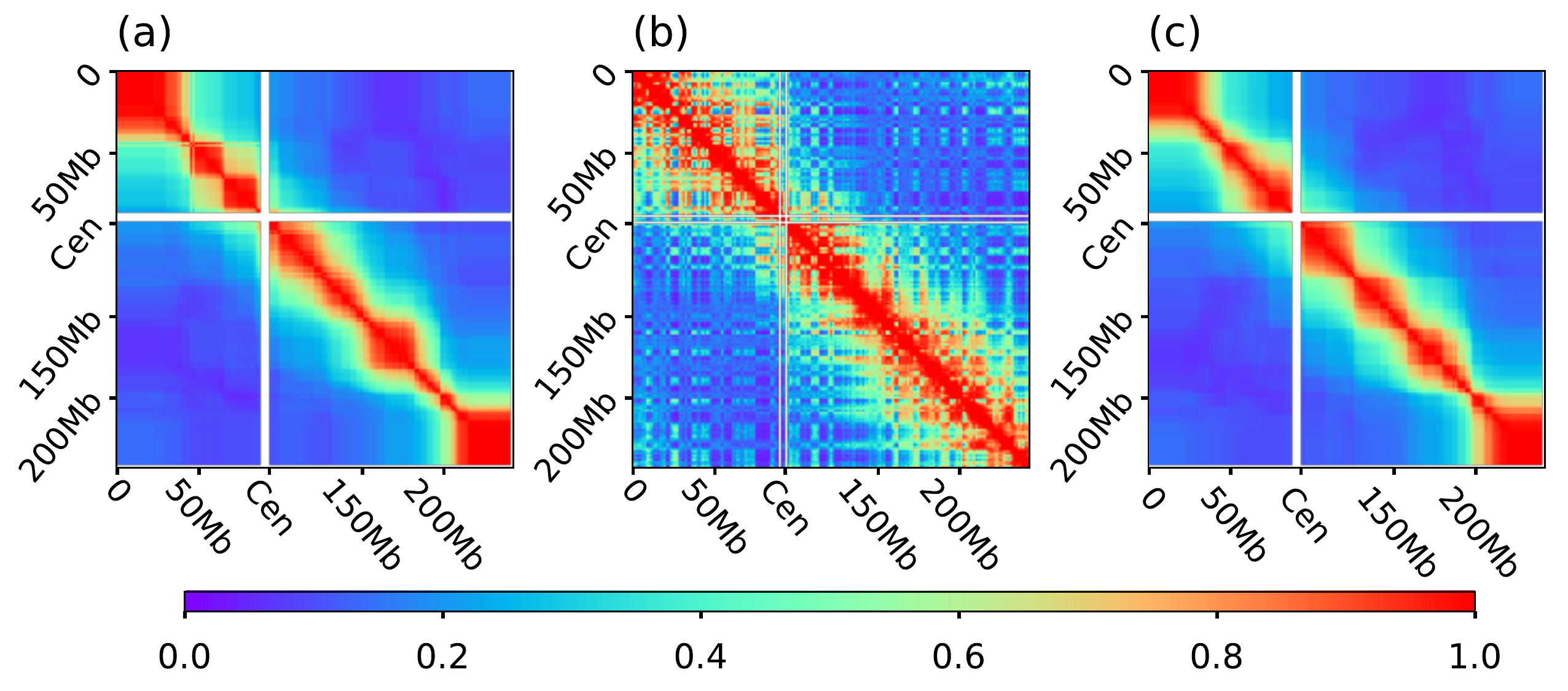}
    \caption{Similarity maps of genomic locations on human chromosome
      2. (a) Averaged consensus matrix $\langle C_\text{LGG} \rangle$
      computed from SCNA data in
      LGG. (b) HiC contact map in normal glial
      cells \cite{Dunham:2012short}. (c) Averaged consensus matrix $\langle C_\text{GBM} \rangle$.  }
    \label{fig:hic}
\end{figure}

For instance, we used the somatic copy number alteration (SCNA) data
in low-grade glioma (LGG) and glioblastoma (GBM) patients from the
Cancer Genome Atlas to construct an adjacency matrix of genomic
locations (SM \cite{Note1}), and performed QTC with the chosen number of
clusters equal to $2, 3, 4,$ or $5$.  
We summarized the predicted
similarity between genomic coordinates by averaging the consensus
matrices $\{C(q)\}_{q=2}^{5}$ for LGG and GBM separately,
 yielding $\langle C_\text{LGG} \rangle$ and
$\langle C_\text{GBM} \rangle$.  
The block structures in SCNA captured by QTC closely resembled the 3D chromatin
interaction HiC contact matrix (Fig.~\ref{fig:hic}) \cite{Dunham:2012short};  the Pearson
correlation coefficients between $\langle C_\text{LGG/GBM} \rangle$
and $\tanh((C_{\text{HiC}})_{ij}/{\bar C_{\text{HiC}}}) \in [0,1)$ was
$0.87$, whereas  the same correlation involving the raw SCNA data was less than $0.50$
(Fig.~S6). Our QTC consensus matrix thus denoises
the SCNA data and helps support the previously observed
phenomenon linking genomic alterations in cancer with the 3D
organization of chromatin \cite{Fudenberg:2011dv}.

In summary, a quantum mechanical wave function is dramatically
different from a classical heat density; even for an initial point
source, the former demonstrates an oscillatory wave behavior, while
the latter is smooth and monotonic in both space and time. Overcoming
the previous difficulties in measuring data similarity using wave
functions, we here devised a stand-alone clustering algorithm based on
quantum transport on network graphs.  Realistic data usually consist
of a large number of features, and the large feature dimensions can
often render clustering algorithms inefficient \cite{Marimont:1979eh}.
Although we do not directly address this issue here, our QTC algorithm
may be combined with known methods for ameliorating the ``curse of
dimensionality'' \cite{Zhao:2017wm}.  Another major challenge in
clustering arises when putative clusters are strongly mixed; in such a
case, supervised learning is usually the most efficient solution by
introducing manually labeled training samples \cite{Hastie:2013fd}.

In addition to high dimensionality and strong mixing, geometric
complexity remains an outstanding challenge; e.g., the cheese-stick
distribution shown in Fig.~\ref{fig:syn_examples}(b) with several
visually separable pieces confuses almost all clustering algorithms.
But, we have demonstrated that the coherent phase information encoded
in the two-point Green's functions, or equivalently the
Laplace-transformed wave functions, are as powerful as the widely
applied spectral clustering. Furthermore, the QTC shows more
robustness when the data distribution contains density fluctuations or
a hierarchy of cluster sizes (SM \cite{Note1}, III A).  Using multiple
initialization sites, QTC generates an ensemble of phase
distributions, which in turn provide a collection of discrete cluster
labels (SM \cite{Note1}, I C). We may either select the most popular
partition from the ensemble or encode the votes from the ensemble
members into a consensus matrix. If most members favor a particular
partition, it is an indication that the clusters are easily separable;
conversely, split votes between several partitions may indicate
suboptimal model parameters or strongly mixed clusters. Thus, QTC
provides a useful self-consistency criterion absent in most clustering
methods. Even in the case of spit votes, the consensus matrix can
still be used in other clustering or supervised learning methods as an
improved similarity measure.  In addition to the consensus matrix, we
have explored other ways of constructing a QT kernel that can be used
as an input to numerous (dis)similarity-based algorithms (SM
\cite{Note1}, III, Fig.~S8 \& S9).  For example, we have tested the
time-average of squared transition amplitude as a similarity measure
in spectral clustering (Fig.~S8 \& S9); the performance was slightly
better than spectral clustering using Gaussian affinity, although some
intrinsic weaknesses of spectral embedding persisted (SM \cite{Note1},
III A).  These results provide evidence for potential benefits that
may arise from studying data science using quantum physics.

We thank Alan Luu,  Mohith Manjunath, and Yi Zhang for their help.
This work was supported by the Sontag Foundation and the Grainger
Engineering Breakthroughs Initiative.

\newpage
\setcounter{figure}{0}    
\setcounter{equation}{0}    

\renewcommand{\thefigure}{S\arabic{figure}}
\renewcommand{\theequation}{S\arabic{equation}}


\centerline{\bf \Large Supplemental Material}

\tableofcontents
\section{Quantum Transport Clustering (QTC)}
\label{sec:qtc}
The Schr\"odinger equation for a free particle
is, up to the Wick rotation $t\rightarrow{\rm i}t$, formally
similar to the heat equation with heat conductance $\kappa$:
\[
\partial_{t}u=\kappa\nabla^{2}u.
\]
Assuming that the heat conductance $\kappa$ is constant in space, the
heat equation can be rewritten as
\[
\partial_{t}u=\kappa\nabla^{2}u=-\nabla\cdot\left(-\kappa\nabla u\right).
\]
Defining the heat current as
\[
\mathbf{j}=-\kappa\nabla u\, ,
\]
the heat equation then becomes the conservation law
\[
\partial_{t}u+\nabla\cdot\mathbf{j}=0.
\]
The Schr\"odinger equation also embodies a conservation law. For example,
consider the Schr\"odinger equation with a time-independent potential $V(x)$:
\[
{\rm
  i}\partial_{t}\psi=-\frac{\nabla^{2}\psi}{2m}+V(x)\psi,
\]
in units where $\hbar=1$. Writing its solution as
$\psi(x,t)=\sqrt{\rho(x,t)}{\rm e}^{{\rm i}\theta(x,t)}$, where $\rho$
is the probability density and $\theta$ the phase, we see that the
Schr\"odinger equation is not one but two coupled equations for $\rho$
and $\theta$,
\[
\dot{\rho}=-\nabla\cdot\left(\rho\frac{\nabla\theta}{m}\right)\equiv-\nabla\cdot\left(\rho\mathbf{v}\right)=-\nabla\cdot\mathbf{j},
\]
where $\mathbf{v}=\nabla\theta/m$ is the group velocity
of a quantum mechanical particle, and $\mathbf{j}=\rho\mathbf{v}$
the current density; and
\begin{align*}
	-\dot{\theta} =&
        \frac{m}{2}\left(\frac{\nabla\theta}{m}\right)^{2}+V-\frac{1}{2m}\left[
        \frac{\nabla^2 \sqrt{\rho}}{\sqrt{\rho}}\right] \\
	\equiv & \frac{1}{2}m\mathbf{v}^{2}+V+Q
\end{align*}
where $Q=-\frac{1}{2m}\left[ \frac{\nabla^2
    \sqrt{\rho}}{\sqrt{\rho}}\right]$ is the ``quantum potential.''

Notice that the quantum current is proportional to $\nabla\theta$
instead of $\nabla\rho$.  Thus, the phase gradient drives the
propagation of the wave function, which encodes richer physics than
classical heat density. This observation suggests that the phase
information may be useful for devising quantum algorithms.


\subsection{Laplace transform of time evolution\label{sec: laplace transform}}
The Laplace transform of a wave function $|\psi(t)\rangle$, evolved
from an initial state $|\psi(0)\rangle$ via a  time-independent
Hamiltonian $H$, is given by
$$
	 |\tilde \psi(s)\rangle \equiv {\cal L}[ |\psi\rangle](s) =
         \int_0^\infty  {\rm e}^{-st} {\rm e}^{-{\rm i}Ht}
         |\psi(0)\rangle\, dt.
$$
Since $H$ is time-independent, we have
$$
	 |\tilde \psi(s)\rangle = \frac{1}{s+{\rm i}H} |\psi(0)\rangle = {\rm i}G({\rm i}s) |\psi(0)\rangle,
$$
where $G(z)\equiv (z-H)^{-1}$ is the resolvent operator of $H$. In the
main text, we interpret $G(z)$ using an effective tight-binding
model. Here, we study the Laplace-transformed wave function
explicitly.  The inverse of the variable $s$ sets the time scale
within which the Schr\"odinger time evolution is averaged; i.e., this
scale sets the extent to which oscillation in time is smoothed out and
destructive interference that can potentially localize the transport
gets ameliorated.  Motivated by this observation, this paper demonstrates that taking the
Laplace transform can resolve the issues of wave function oscillation
and localization that have hindered the application of quantum
mechanics to clustering problems.

Of note, recall that spectral clustering uses the $j$-th entries of
the first few lowest-eigenvalue eigenvectors of the graph Laplacian to
represent the $j$-th node.  By contrast, one distinct advantage of QTC
lies in utilizing the eigenvectors $\psi_n$ twice when computing the
phase of
$$
	 \langle j |\tilde \psi(s)\rangle =   \sum_n \frac{\langle
           \psi_n | \psi(0) \rangle}{s+{\rm i}E_n} \psi_n(j);
$$
namely, both the $j$-th entries $\psi_n(j)$, just as in spectral
clustering, and the projections $\langle \psi_n | \psi(0) \rangle $
onto the initialization node are used.  In this way, as the
intialization node varies during the random sampling step, the phase
representations of two nodes within a cluster will stay close to each
other, and this information is pooled together in the QTC algorithm.


\subsection{Choosing the number of clusters\label{sec:n_cluster}}
\begin{figure}[h]
    \centering
    \includegraphics[width=1.0\columnwidth]{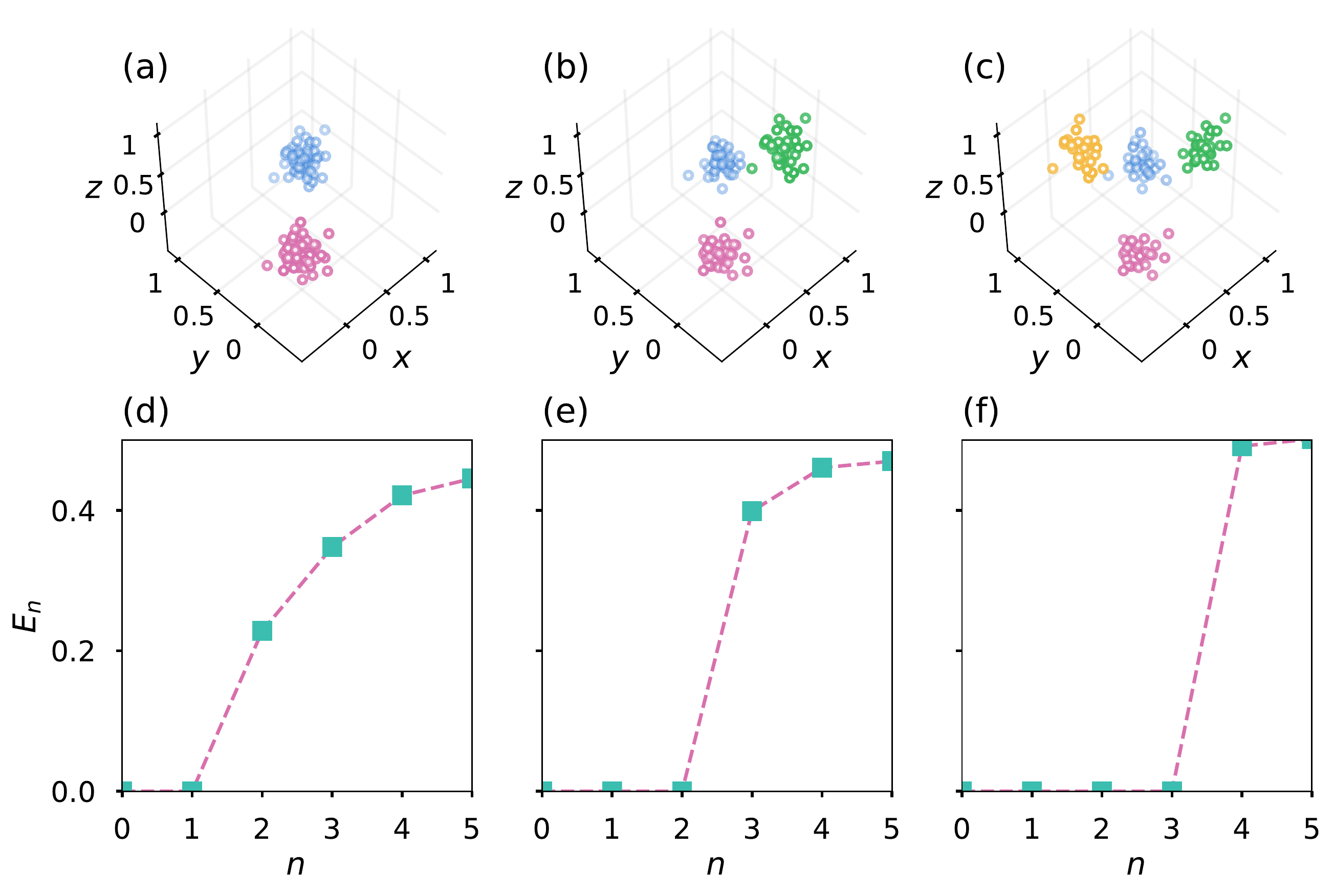}
    \caption{(a,b,c) Gaussian distributions ($\sigma=0.1$) in $\mathbb
      R^3$, with the means located at the vertices of a regular
      tetrahedron of length 1. The inter-cluster distance is thus
      $10\sigma$.  (d,e,f) The spectrum of symmetric normalized graph
      Laplacian $H$ corresponding to the data distributions in
      (a,b,c), respectively.}
    \label{fig:n_cluster}
\end{figure}

If $q>1$ clusters are well-separated, the Hamiltonian is approximately
$q$-block diagonal.  Fluctuations between the $q$ macroscopic modes
have lower kinetic energy, which mainly arises from inter-cluster
tunneling, than microscopic fluctuations within each cluster.  In this
case, there exists an energy gap separating the low-energy macroscopic
modes from the high-energy microscopic oscillations.  Furthermore, the
low-energy states can be approximated as linear combinations of
cluster wave functions; thus, the number of low-energy states equals
the number of putative clusters.  For illustration, we generated
well-separated $q = 2, 3, \text{and } 4$ Gaussian clusters in three
dimensions (Fig.~\ref{fig:n_cluster}(a,b,c)); the adjacency matrix was
computed using the $10\%$-quantile of pairwise distance distribution
as the proximity scale in Gaussian kernel. The first 6 eigenvalues of
the Hamiltonian are plotted in Fig.~\ref{fig:n_cluster}(d,e,f).

\subsection{Phase information\label{sec:phase_info}}
In applications, we numerically calculate the Laplace transform of a
wave function initialized at a given node and then extract the phase
distribution. As in the main text, we will assume that the total number of
nodes is $m$ and the {\em a priori} determined number of clusters is
$q$.
 The phases of nodes belonging to different clusters are
typically separated by gaps, allowing us to assign discrete
class labels to nodes. We propose two methods for  converting the phases  to
class labels $0,1,\ldots, q-1$: (Method 1) direct difference, and
(Method 2) clustering.  The steps in Method
1 are as follows:

\begin{enumerate}
  \item[] \hspace{-4mm}\underline{\bf Method 1}
	\item   Sort the array $(\theta_0,\ldots,\theta_{m-1})$ of
          phases in ascending order. Let $\pi(i)$ denote the rank
          of the phase of node $i$ in this sorted list. 
	\item Denote the
          $j$-th element in the sorted list as $\theta_{(j)}$ and compute $\hat n_j = (\cos \theta_{(j)}, \sin
          \theta_{(j)})^\top \in \mathbb{R}^2,$ for $j=0,\ldots,m-1$.
	\item Compute the local difference $r_{j} = \| \hat n_{j+1} - \hat
          n_{j} \|$, for $j=0,1,\ldots,m-2$ \footnote{We did not use arc length, or the geodesic distance on $S^1$ in that arc length is sensitive to fluctuations in phase distribution; however, our goal is to pick out the largest jumps in phases and ignore small jumps which may arise around large jumps.}.
	\item  Locate the $q-1$ largest values in the array $(r_0,\ldots,r_{m-2})$ and
          return their indices $\{I_j\}_{j=1}^{q-1}$, where $I_j<I_{j+1}$.
	\item Assign the class label $j$ to node $i$ iff $I_{j}<\pi(i) \leq I_{j+1}$, where $I_0=-1$ and $I_{q} =m-1$.
\end{enumerate}
The steps in Method 2 are as follows:
\begin{enumerate}
  \item[]\hspace{-4mm} \underline{\bf Method 2}
	\item Map each node $i$ to $\hat n_i = (\cos \theta_i, \sin \theta_i)^\top\in\mathbb{R}^2$.
	\item Apply a standard clustering algorithm in $\mathbb{R}^2$, e.g. $k$-means
          or $k$-medoids.
	\item Return the class label for each node
\end{enumerate}

The first method is faster than the second method. However, when the
clusters are not clearly separable it might recognize false cluster
boundaries and produce fragmented clustering. We find that the second method  is more robust.

Using either Method 1  or Method 2, we are thus
able to convert the phase distribution of  a Laplace transformed wave function
initialized at a single node to a set of
discrete class labels.  When we change the intialization node, some of
the cluster boundaries can change. To improve clustering accuracy and reduce variation
in clustering, we thus iterate QTC at multiple nodes; let $m'$ denote
this number of initialization nodes. The
clustering results then form an ensemble of class labels, organized
into a matrix
$(\Omega_{ij})$, where $i=0,1,\ldots,m-1$ runs through all nodes and
$j=0,1,\ldots,m'-1$ indexes the iteration of initialization. 

Notice that the
class labels may get permuted across different initialization. 
We introduce two methods to handle this issue and  summarize the $\Omega$-matrix: (1)
direct extraction, and (2) consensus matrix.


\subsubsection{Direct extraction \label{sec:algo_dir_extr}}
We want to count the multiplicity of the columns of
$\Omega$, up to permutation of class labels; i.e.~two columns are
considered equivalent if they are equal upon permuting the class
labels.  We will then choose the most
frequent column vector as the desired partition of nodes.
For this purpose, we first devise a scheme for testing whether a
subset of columns are all equivalent.
 
Let $\{p_{i}\}=\{2,3,5,7,\cdots\}$ be the set of primes, then
$\{\sqrt{p_{i}}\}$ is a set of irrational numbers serving as linearly
independent  vectors over the field $\mathbb Q$ of rational numbers.
Let $A$ be an index set containing at least two column indices of
$\Omega$. For each node $i$, we
then compute the quantity
$\xi_i=\sum_{k\in{A}}\Omega_{i k}\sqrt{p_{k}}$.  For any two nodes $i$ and $j$,
\begin{equation}
	\xi_{i}-\xi_{j}=\sum_{k\in A}\left(\Omega_{ik}-\Omega_{jk}\right)\sqrt{p_{k}}\equiv\sum_{k\in A}b_{k}\sqrt{p_{k}}.
\end{equation}
Suppose $i$ and $j$ are in the same cluster for all $k\in A$, then
$b_{k}=0$ for all $k$, and thus $\xi_{i}=\xi_{j}$; the converse is
also true, because $\{\sqrt{p_{i}}\}$ are linearly independent over
$\mathbb Q$.  Thus, $\xi_i=\xi_j$ iff node $i$ and  node $j$ are
assigned to the same class by all columns indexed by $A$. The minimum number
of distinct $\xi_i$ is $q$, since any column of $\Omega$ partitions the
nodes into $q$ clusters. If the number of distinct $\xi_i$ exceeds
$q$, then there thus exists at least two columns that disagree on the
partition, so the columns indexed by $A$ are not all equivalent.

Our algorithm including this scheme is as follows:
\begin{enumerate}
\item[] \hspace{-4mm}\underline{\bf Ensemble Method 1}

\item Let $K=\{0,1,\cdots, m'-1\}$ be the full index set indexing the
  columns of $\Omega$. Denote any non-empty subset of $K$ as $K'$, and
  let $k'_0$ denote the first column index appearing in $K'$.

\item \textbf{Define} function ${\sf IsEquiv}(\{\Omega_{ik}\}_{k\in
    K'})$ to tell whether the columns of $\Omega$ indexed by $K'$
  yield an \emph{equivalent clustering}:
\begin{itemize}
\item[] \textbf{For} $i=0,1,\ldots,m-1$:
	\begin{itemize}
		\item[] $\quad\quad\xi_{i}=\sum_{k\in K'}\Omega_{ik}\sqrt{p_{k}}$
	\end{itemize}

\item[] Count the number $q'$ of distinct $\xi_i$
\item[] \textbf{If} $q'=q$, then \textbf{Return} {\sf 
True}
\item[] \textbf{Else}: \textbf{Return} {\sf False}
\end{itemize}

\item Let $\sf H$ be a hash table with non-negative integer keys
  $\alpha$ indexing the equivalence classes of columns of $\Omega$ and
  values ${\sf H}_\alpha$ equal to the corresponding index sets of
  equivalent columns. Each key $\alpha$ is chosen from ${\sf
      H}_\alpha$ to represent the class.
\item \textbf{Define} function ${\sf Pigeonhole}(\{\Omega_{ik}\}_{k\in K'},\:{\sf H})$:
\begin{itemize}
\item[] \textbf{If} ${\sf IsEquiv}(\{\Omega_{ik}\}_{k\in K'})={\sf True}$,
then:
\begin{itemize}
\item[] ${\sf IsExisting=False}$
\item[] \textbf{For} $\alpha$ in ${\sf H}$:
\begin{itemize}
\item[] \textbf{If} ${\sf IsEquiv}(\{\Omega_{ik}\}_{k=\alpha, k'_0})={\sf True}$: 
\begin{itemize}
\item[] ${\sf IsExisting=True}$
\item[] Merge $K'$ and ${\sf H}_{\alpha}$
\item[] \textbf{break} for-loop
\end{itemize}
\end{itemize}
\item[] \textbf{If} ${\sf IsExisting=False}$:
\begin{itemize}
\item[] Create a new key $\alpha'$ and ${\sf H}_{\alpha'}=K'$
\end{itemize}
\end{itemize}
\item[] \textbf{Else}: Split $K'$ in two halves, $K'_{1}$ and $K'_{2}$
\begin{itemize}
\item[] \textbf{Call} ${\sf H}={\sf Pigeonhole}(\{\Omega_{ik}\}_{k\in K'_{1}},\:{\sf H})$
\item[] \textbf{Call} ${\sf H}={\sf Pigeonhole}(\{\Omega_{ik}\}_{k\in K'_{2}},\:{\sf H})$
\end{itemize}
\item[] \textbf{Return} \emph{${\sf H}$}
\end{itemize}
\item \textbf{Call} ${\sf Pigeonhole}(\{\Omega_{ik}\}_{k\in K},\:{\sf H}^{0})$,
where ${\sf H}^{0}$ is an empty hash table
\end{enumerate}


\subsubsection{Consensus matrix \label{sec:algo_consencus}}
Even though the class labels may get randomly permuted for different
initializations, whether two nodes
share the same class label within each initialization is independent
of the labeling convention.
Therefore, we define a
consensus matrix $C$ with elements
\begin{equation}
	C_{ij} = \frac{\sum_{k=1}^{m'} \delta(\Omega_{ik} - \Omega_{jk})}{m'},
\end{equation}
where $\delta$ is the Kroneker delta or indicator function, and $m'
\le m$ is the number of the chosen initialization nodes.  Notice that
$C_{ij}=C_{ji} \in [0,1]$, and $C_{ii}=1$ for all nodes
$i,j=1,2,\ldots,m$. The algorithm is sketched as follows:
\begin{enumerate}
\item[] \hspace{-4mm}\underline{\bf Ensemble Method 2}
\item Initialize $C$ as an $m\times m$ identity matrix
\item \textbf{For} $i=0,1,\ldots,m-1$:
\begin{itemize}
\item[] \textbf{For} $j=i+1,\ldots,m-1$:
\begin{itemize}
\item[] \textbf{For} $k=0,1,\ldots,m'-1$:
\begin{itemize}
\item[] \textbf{If} $\Omega_{ik}=\Omega_{jk}$: $C_{ij}++$
\end{itemize}
\item[] $C_{ji}=C_{ij}$
\end{itemize}
\end{itemize}
\item $C_{ij}=C_{ij}/m'$ for $i\neq j$
\end{enumerate}
The consensus matrix measures the similarity of node pairs and
facilitates the visualization of network structure, e.g.~chromatin
interaction information
between distal genomic loci, as in Fig.~4.  It can also be used as a
similarity measure or dissimilarity measure,
e.g.~$\delta_{ij}-C_{ij}$, in (dis)similarity-based algorithms such as
spectral clustering and hierarchical clustering.

\subsection{Effective tight-binding model\label{sec:eff_theory}}
In the extreme case where the clusters are completely separated from
each other, the Hamiltonian $H$ is strictly in $q$ diagonal blocks;
each block governs the dynamics within a cluster and has its own
ground state wave function $\phi_\mu (i) = \langle i | \phi_\mu
\rangle$, which is positive for node $i$ belonging to the $\mu$-th
cluster and zero otherwise. We have $H |\phi_\mu\rangle = \xi_\mu |\phi_\mu
\rangle$ and $\langle \phi_\mu | \phi_\nu \rangle = \delta_{\mu\nu}$
for all $\mu,\nu=0,1,\ldots,q-1$.  As we gradually turn on
off-diagonal couplings $v_{\mu\nu} = \langle \phi_\mu | H | \phi_\nu
\rangle$ between clusters $\mu \ne \nu$, the wave functions $\phi_\mu$
are no longer eigenstates of $H$. The effective tight-binding model
assumes that in the weak coupling limit, we can project $H$ onto the
subspace spanned by $\{\phi_\mu\}_{\mu=0}^{q-1}$ and diagonalize the
projected Hamiltonian $h_{\mu\nu} = \langle \phi_\mu | H | \phi_\nu \rangle$
to approximate the first $q$ lowest energy eigenstates.

The resolvent matrix $g_{\mu\nu}$ of $h_{\mu\nu}$ is defined through
$$
g^{-1}(z)_{\mu\nu} = z\delta_{\mu\nu}-h_{\mu\nu}. 
$$
The resolvent matrix can  be expanded if $|v_{\mu\nu}| < |z-\xi_\nu|$,
for all $\mu,\nu = 0, 1, \ldots, q-1$, as
\begin{widetext}
\begin{equation}
  g_{\mu\nu}(z)=\frac{\delta_{\mu\nu}}{z-\xi_{\mu}}+\frac{v_{\mu\nu}}{(z-\xi_{\mu})(z-\xi_{\nu})}+\sum_{\sigma}\frac{v_{\mu\sigma}v_{\sigma\nu}}{(z-\xi_{\mu})(z-\xi_{\sigma})(z-\xi_{\nu})}+\sum_{\sigma,\rho}\frac{v_{\mu\sigma}v_{\sigma\rho}v_{\rho\nu}}{(z-\xi_{\mu})(z-\xi_{\sigma})(z-\xi_{\rho})(z-\xi_{\nu})}+{\cal O}(v^{4}).
\end{equation}	
\end{widetext}
Note that the resolvent matrix is thus a weighted sum over all
possible tunneling paths between the $q$ clusters.

\begin{figure}[t]
    \centering
    \includegraphics[width=1.0\columnwidth]{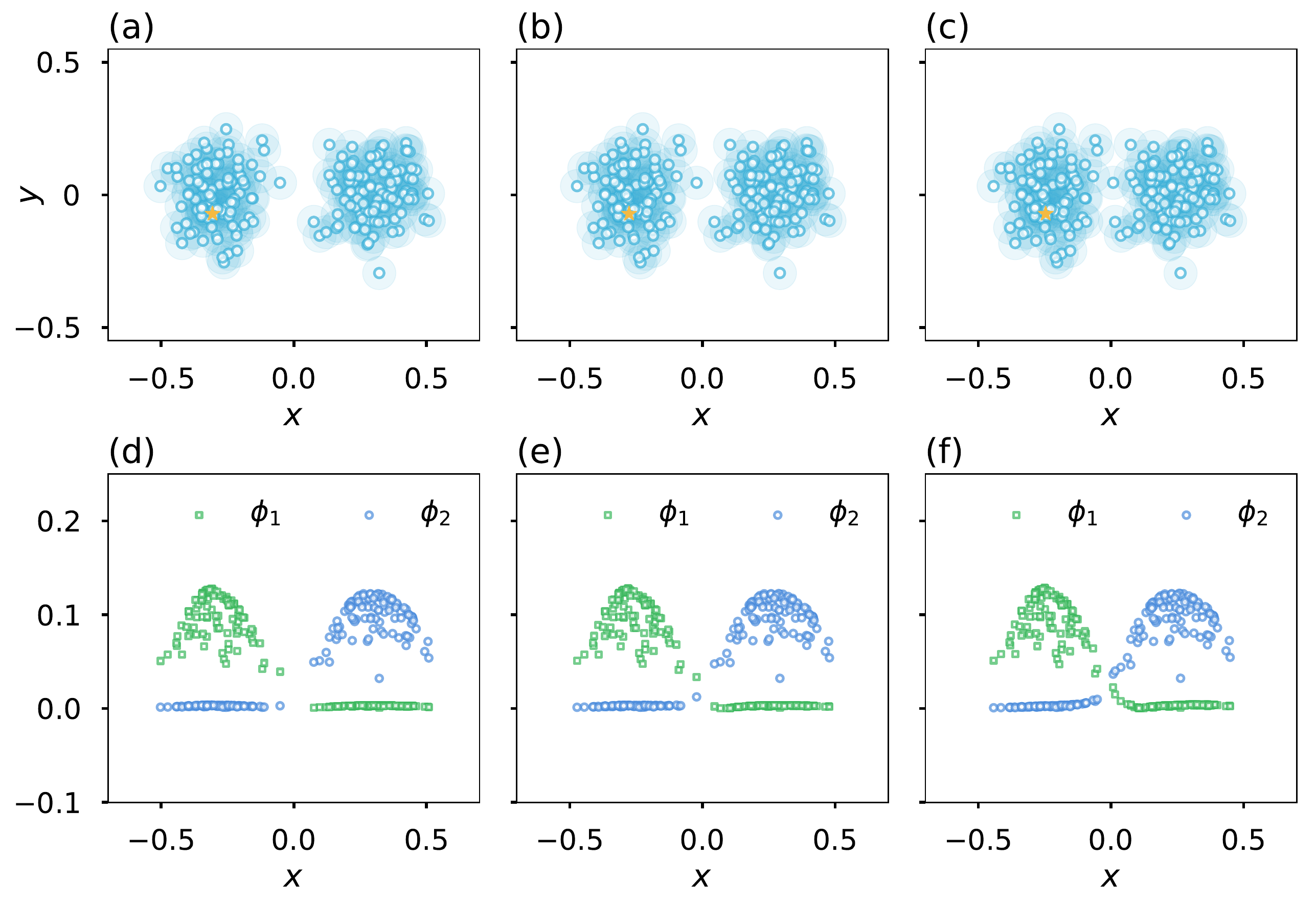}
    \caption{(a-c) Two-cloud distributions corresponding to
      Fig.~3(a-c). (d-f) Cluster wave functions used to compute
      the theoretical predictions in Fig.~3(d-f).}
    \label{fig:eff_ham_wf}
\end{figure}
\subsection{Two-level toy model \label{sec:two_level}}
\begin{figure}
    \centering
    \includegraphics[width=1.0\columnwidth]{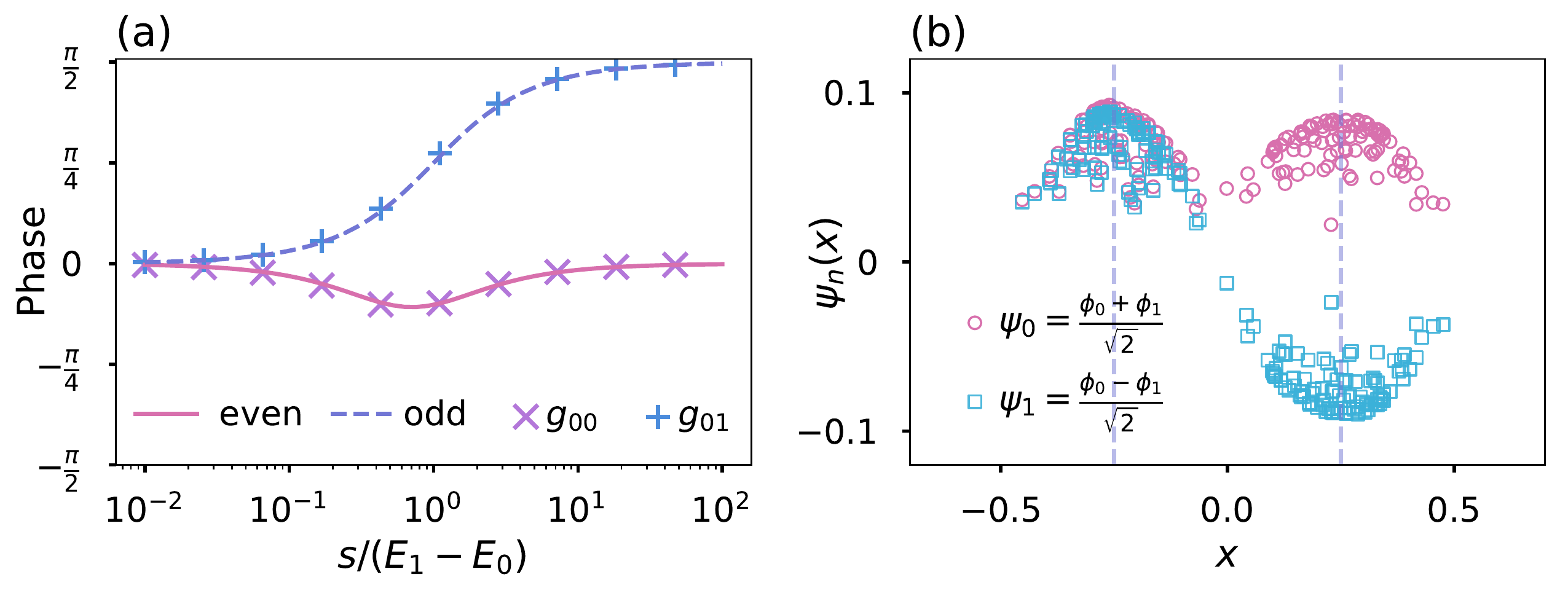}
    \caption{(a) The phase distribution of the Laplace transform of exact
      instanton solution (solid and dashed lines represent $G_{00}$
      and $G_{01}$, respectively). Also plotted are the phases calculated from our
      two simulated Gaussian clouds ${\cal N}((\pm \ell, 0)^\top,
      \sigma^2{\bf 1}_{2\times 2} )$, with $\ell = 0.25$, $\sigma=0.1$,
      and equal sample size $m=100$ ($\times$ and $+$). (b) Plots of the
      ground state $\psi_0$ and the first excited state $\psi_1$ wave
      functions derived from the simulated data.}
    \label{fig:1.instanton}
\end{figure}

Consider the case of two Gaussian clusters in $\mathbb R^2$ with mean
at $(\pm\ell, 0)^\top$, as shown in Fig.~3(a-c) and
Fig.~\ref{fig:eff_ham_wf}(a-c).  We expect two low energy states,
i.e.~the ground state and the first excited state
(Fig.~\ref{fig:1.instanton}(b)).  Let $\phi_0$ and $\phi_1$ denote the
cluster wave functions for the left and right Gaussian clouds,
respectively. Assuming that the two clusters have the same ground
state energy, the ground state $\psi_0$ and the first excited state
$\psi_1$ of the tight-binding Hamiltonian are
$$
  |\psi_0\rangle = \frac{|\phi_0\rangle + |\phi_1\rangle}{\sqrt
    2},\quad |\psi_1\rangle = \frac{|\phi_0\rangle -
    |\phi_1\rangle}{\sqrt 2}. 
$$

Setting the ground state energy $E_{0}=0$, and defining the first
energy gap
$E\equiv E_{1}-E_{0}$, we have
\begin{equation}
  |\tilde\psi(s)\rangle=\frac{1}{s+{\rm i}H}|\psi(0)\rangle \approx
  \frac{c_{0}|\psi_{0}\rangle}{s}+\frac{c_{1}|\psi_{1}\rangle}{s+{\rm
      i}E}, 
\end{equation}
where $c_j = \langle \psi_j | \psi(0) \rangle$. 
Thus,
$$
	|\psi(s) \rangle =
        \frac{\left(\frac{c_{0}}{s}+\frac{c_{1}}{s+{\rm
                i}E}\right)|\phi_{0}\rangle +
          \left(\frac{c_{0}}{s}-\frac{c_{1}}{s+{\rm
                i}E}\right)|\phi_{1} \rangle}{\sqrt 2}, 
$$
from which we easily extract the phase in the left and right clusters
to be
\begin{align*}
	\Theta_0 &= \arg \left(\frac{c_{0}}{s}+\frac{c_{1}}{s+{\rm i}E}\right)\notag\\
	&=\arctan\frac{Ec_{0}}{(c_{0}+c_{1})s}-\arctan\frac{E}{s},
        \text{ and}\\
	 \Theta_1 &= \arg \left(\frac{c_{0}}{s}-\frac{c_{1}}{s+{\rm i}E}\right)\notag\\
	 &=\arctan\frac{Ec_{0}}{(c_{0}-c_{1})s}-\arctan\frac{E}{s}.
\end{align*}
If the initial state $\psi(0)$ is a delta function located deep in the (1)
left or (2) right cluster, then (1) $c_0 = c_1$ or (2) $c_0 = -c_1$,
respectively.  The phases of the left and right clusters in case (1)
are
\begin{align}\label{eq:Theta}
	\Theta_{00} &= \arctan\frac{E}{2s} - \arctan\frac{E}{s}\notag\\
	\Theta_{01} &= \frac{\pi}{2} - \arctan\frac{E}{s};
\end{align}
while in case (2), the phases are
\begin{align}
	\Theta_{10} &= \frac{\pi}{2} - \arctan\frac{E}{s}\notag\\
	\Theta_{11} &= \arctan\frac{E}{2s} - \arctan\frac{E}{s}.
\end{align}
Notice that $\Theta_{\mu\nu}$ is a constant diagonal symmetric matrix
that preserves the left-right symmetry.

The two-cluster model can be mapped to the classic double-well
instanton tunneling model which will be briefly summarized below;
detailed derivations can be found in \cite{Novikov:1999gs}. The model
Hamiltonian is
\[
H=-\frac{1}{2}\partial_{x}^{2}+\lambda(x^{2}-\ell^{2})^{2},
\]
where $\lambda >0$.  The potential $V(x)=\lambda(x^{2}-\ell^{2})^{2}$
has two minima at $x=\pm\ell$ for $\ell>0$ and one minimum at $x=0$
for $\ell=0$. The barrier height is $V(0)=\lambda\ell^{4}$ which grows
rapidly with the separation distance $\ell$. In the vicinity of
minima,
$V(\pm\ell+\varepsilon)=\lambda(\pm2\varepsilon\ell+\varepsilon^{2})^{2}=4\lambda\ell^{2}\varepsilon^{2}+{\cal
  O}(\varepsilon^{3})$; the local harmonic frequency is thus $\omega =
2\ell\sqrt{2\lambda}$ and $V(0)=\omega^{4}/64\lambda$.

In the limit $\lambda\downarrow0$ while keeping $\omega$
constant, the barrier is infinite,  and the ground state is two-fold degenerate with harmonic
ground state energy $E_{0}=\frac{1}{2}\omega$ and expected position
$\langle x\rangle=\pm\ell$. For any finite barrier, however, we should
have $\langle x\rangle=0$, which is enforced by symmetry; the symmetric
solution cannot be obtained via perturbation around either of the
local minima.

Non-perturbative instanton solution splits the degeneracy: 
\begin{align*}
E_{0} & =\frac{\omega}{2}\left(1-2\sqrt{\frac{\omega^{3}}{2\pi\lambda}}{\rm e}^{-\omega^{3}/12\lambda}\right),\\
E_{1} & =\frac{\omega}{2}\left(1+2\sqrt{\frac{\omega^{3}}{2\pi\lambda}}{\rm e}^{-\omega^{3}/12\lambda}\right).
\end{align*}
The transition amplitudes are
\begin{align*}
\langle+\ell|{\rm e}^{-{\rm i}Ht}|-\ell\rangle & ={\rm
  i}\sqrt{\frac{\omega}{\pi}}{\rm e}^{-{\rm i}\omega t/2}\sin(\omega
\rho_{\text{inst}}t) \ \text{ and}\\
\langle-\ell|{\rm e}^{-{\rm i}Ht}|-\ell\rangle &
=\sqrt{\frac{\omega}{\pi}}{\rm e}^{-{\rm i}\omega t/2}\cos(\omega
\rho_{\text{inst}}t), 
\end{align*}
where the instanton density $\rho_{\text{inst}}=\sqrt{\frac{\omega^{3}}{2\pi\lambda}}{\rm e}^{-\omega^{3}/12\lambda}$.
Notice that the energy gap is $E=2\omega\rho_{\text{inst}}$; thus,
\begin{align}\label{eq:instantonAmplitude}
\langle\pm\ell|{\rm e}^{-{\rm i}Ht}|-\ell\rangle & =\sqrt{\frac{\omega}{\pi}}{\rm e}^{-{\rm i}\omega t/2}\frac{{\rm e}^{{\rm i}Et/2}\mp{\rm e}^{-{\rm i}Et/2}}{2}\notag\\
 & =\sqrt{\frac{\omega}{\pi}}\frac{{\rm e}^{-{\rm i}E_{0}t}\mp{\rm e}^{-{\rm i}E_{1}t}}{2}\notag\\
 & =\sqrt{\frac{\omega}{\pi}}{\rm e}^{-{\rm i}E_{0}t}\frac{1\mp{\rm e}^{-{\rm i}Et}}{2}.
\end{align}
If we reset the ground state energy to zero, the Laplace transform of
Eq.~\ref{eq:instantonAmplitude} yields the resolvent matrix elements
\begin{align}
g_{00}({\rm i}s) &
=\frac{1}{2}\sqrt{\frac{\omega}{\pi}}\left(\frac{1}{s}+\frac{1}{s+{\rm
      i}E}\right)
\ \text{ and}\notag\\
g_{01}({\rm i}s) & =\frac{1}{2}\sqrt{\frac{\omega}{\pi}}\left(\frac{1}{s}-\frac{1}{s+{\rm i}E}\right),
\end{align}
where 0 and 1 denote the states localized at $x=-\ell$ and  $x=+\ell$, respectively.
The phases are thus
\begin{align}
\Theta_{00}(s) & =\arctan\frac{E}{2s}-\arctan\frac{E}{s},\notag\\
\Theta_{01}(s) & =\frac{\pi}{2}-\arctan\frac{E}{s}.
\end{align}
Note that the above phase distribution is exactly the same as that from the
low-energy two-cluster model (Eq.~\ref{eq:Theta}) upon identifying the energy gaps.

The phase separation between the diagonal and off-diagonal elements of
the resolvent is $\pi/2 -\arctan\frac{E}{2s}$, and this difference is
thus controlled by the ratio $s/E$. In other words, the Laplace
transform parameter $s$ controls the separability between clusters in
the QTC algorithm.  For $s \ll E$, $s=E/2$, or $s \gg E$, the phase
differences are $0$, $\pi/4$, or $\pi/2$,
respectively. Fig.~\ref{fig:1.instanton}(a) shows the phases
$\Theta_{00}$ and $\Theta_{01}$ for different values of $s/E$ in the
range $[10^{-2}, 10^2]$, suggesting that $s$ should be chosen to be at
least as large as the energy gap $E$.

\begin{figure}[h]
    \centering
    \includegraphics[width=1.0\columnwidth]{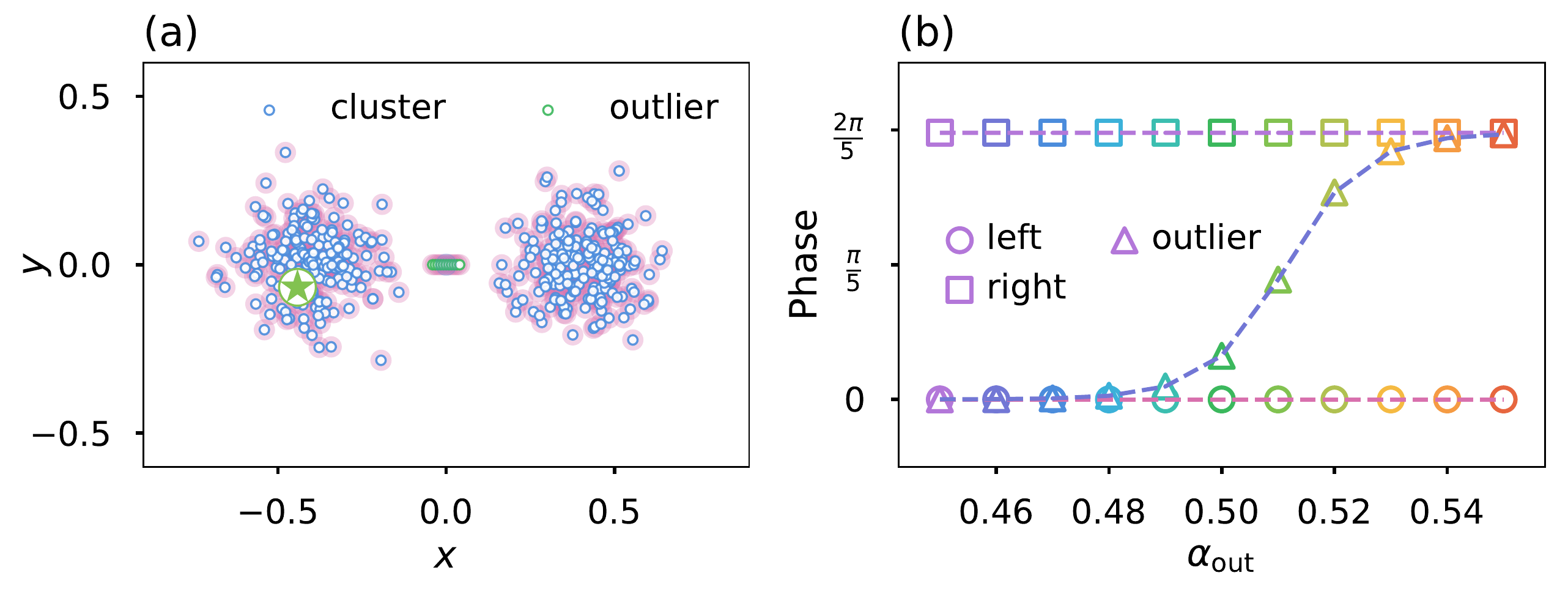}
    \caption{(a) Two Gaussian clusters were drawn from ${\cal N}((\pm
      \ell, 0)^\top, \sigma^2{\bf 1}_{2\times 2} )$ with $\sigma=0.1$,
      sample size $m=100$, and $\ell=0.4$ chosen to yield proximity $r_{5\%}\approx \sigma$; the
      outlier was located at $( -\ell(1-\alpha_\text{out}) +
      \ell\alpha_\text{out} ), 0)^\top$ between the two clusters. (b)
      The quantum transport was initialized from a node in the left
      cluster (marked with $\star$). The phases of the left and
      right clusters, averaged over their respective nodes,  and the
      phase of the outlier are plotted against
      $\alpha_\text{out}$, with the left cluster phases set to zero.}
    \label{fig:outlier_c}
\end{figure}

\begin{figure}[h]
    \centering
    \includegraphics[width=1.0\columnwidth]{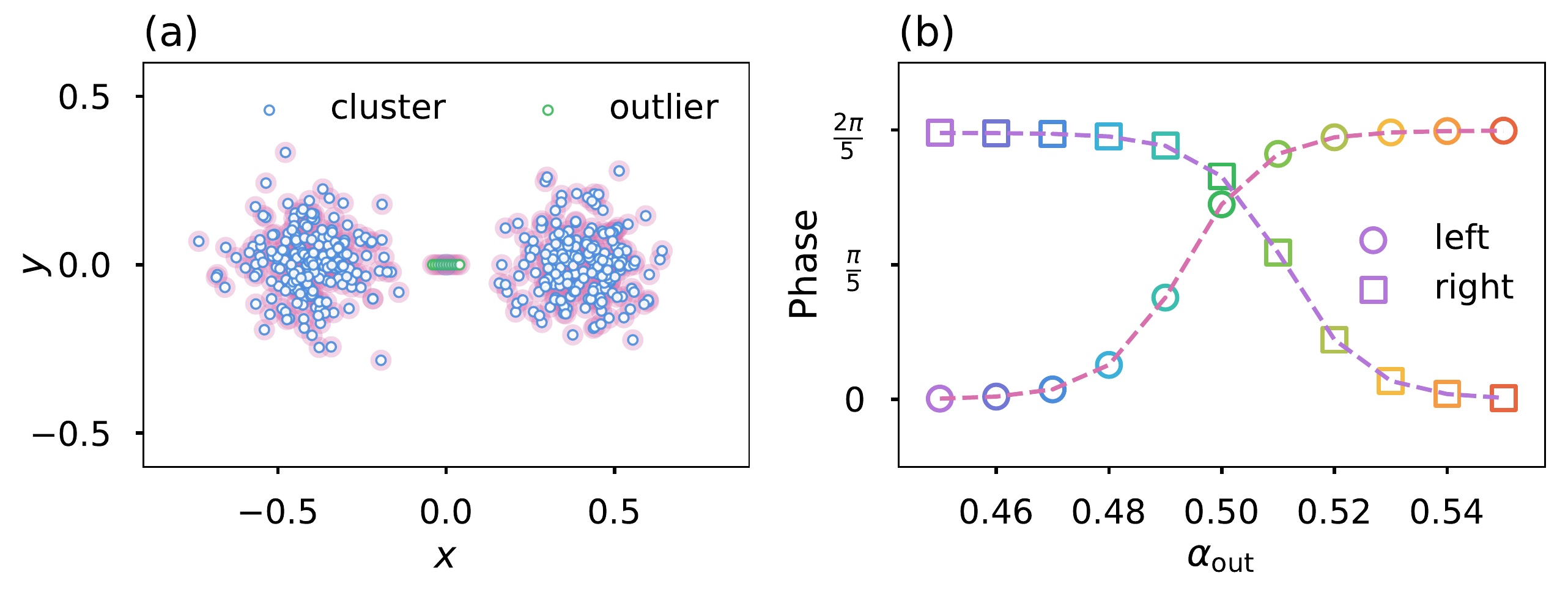}
    \caption{(a) Two Gaussian clusters were drawn from ${\cal N}((\pm
      \ell, 0)^\top, \sigma^2{\bf 1}_{2\times 2} )$ with $\sigma=0.1$,
      sample size $m=100$, and $\ell=0.4$ chosen to yield proximity $r_{5\%}\approx \sigma$; the
      outlier was located at $( -\ell(1-\alpha_\text{out}) +
      \ell\alpha_\text{out} ), 0)^\top$ between the two clusters. (b)
      The quantum transport was initialized from the outlier, and the
      averaged phases of the left and right clusters are plotted against
      $\alpha_\text{out}$.}
    \label{fig:outlier_o}
\end{figure}

In practice, for an ambiguous point located between two clusters, its
phase interpolates smoothly between the cluster phases.  Figure
\ref{fig:outlier_c}(b) shows the phases of the outlier for QTC
initialized from a point deep in the left cluster.  Moreover, Figure
\ref{fig:outlier_o}(b) shows the mean phases of the left and right
clusters for QTC initialized at an outlier located at $(
-\ell(1-\alpha_\text{out}) + \ell\alpha_\text{out} ), 0)^\top$, and it
demonstrates that a wave function initialized from an ambiguous point
loses contrast between the two clusters.

Similarly, for cases involving more than two clusters, the full
$\Theta$-matrix for all nodes essentially amounts to the effective
tight-binding matrix $\arg({\rm i} g_{\mu\nu}({\rm i}s))$.  Our
experience shows that choosing $s$ based on the average gap, $E =
(E_{q-1} - E_0)/(q-1)$, still provides a helpful guideline and yields
good multiclass clustering results.

%

\section{Data Preparation\label{sec:data_prep}}

\subsection{Synthetic Data Sets}
For a sufficiently small proximity measure $r_\varepsilon$ in the
synthetic data in Fig.~1 (b-d \& f-h), both QTC and spectral
clustering were able to produce the correct clustering results. But,
as $r_\varepsilon$ increased, spectral clustering made mistakes, while
QTC remained robust. For sufficiently large proximity values, both spectral
clustering and QTC failed to recognize the clusters.  Thus, there was
a finite interval of $\varepsilon$ for each data set in which QTC
outperformed spectral clustering.  For the data sets in Fig.~1 (b-d \&
f-h), the intervals were approximately $[3.1\%, 3.9\%]$, $[0.61\%,
0.85\%]$, and $[0.39\%, 0.46\%]$, respectively.

\subsection{Time Series Stock Price Data\label{sec:stock}}
The stock price data consisted of the ``adjusted close'' prices of the
AAPL and GOOGL stocks between January 3, 2005 and November 7, 2017,
downloaded from Yahoo Finance. We $\log$ transformed the data and
subtracted the two time series by the respective $\log$-prices on the
first day (1-3-2005).  We computed the pairwise Euclidean distance in
$\mathbb R^2$ and took $1\%$-quantile of the distance distribution as
the proximity length $r_{1\%}=0.05$. Next, we assembled the Gaussian
similarity measure $A_{ij} = \exp[-(r_{ij}/r_{1\%})^2]$ and performed
QTC and spectral clustering; the number of clusters was chosen to be
five.  Spectral clustering was able to produce the clustering
  obtained by QTC at $1\%$-quantile only for shorter proximity lengths
  $\varepsilon \in [0.2\% , 0.5 \%]$; for $\varepsilon \lesssim
  0.1\%$, the clusters started to become disjoint subnetworks.


\subsection{Genomic Data\label{sec:scna}}
\begin{figure}[h]
    \centering
    \includegraphics[width=1.0\columnwidth]{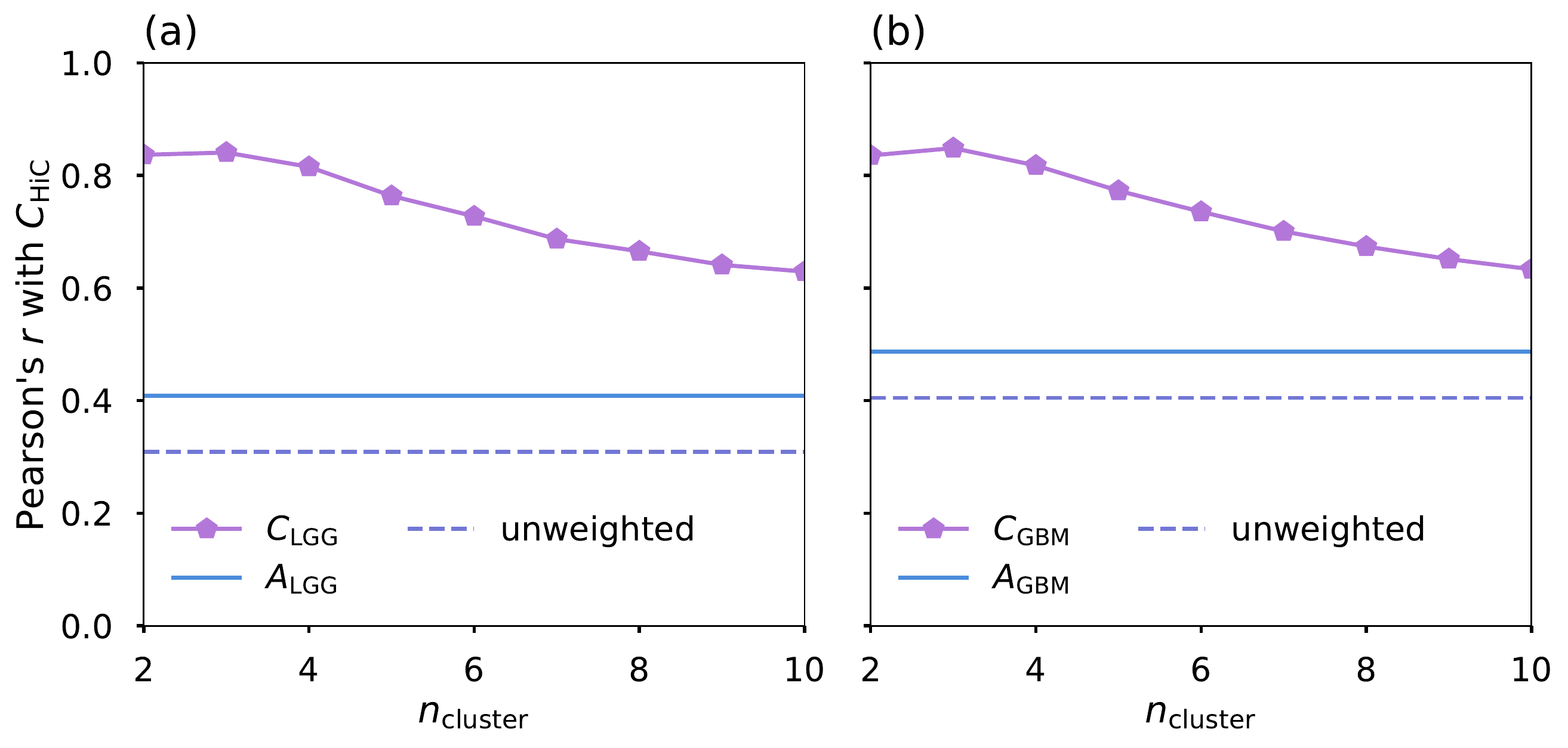}
    \caption{Pearson correlation coefficients between the
      $\tanh$-normalized HiC matrix and various similarity
      measures. For (a) LGG and (b) GBM samples, respectively,
      correlations were computed using the
      ``unweighted''  raw counts $N_{ij}$ of SCNA labeled by
      genomic location pair $(i,j)$, the weighted adjacency
      $(A_\text{LGG/GBM})_{ij}=N_{ij}w_{ij}$ with Gaussian weight
      $w_{ij}=\exp(-(r_{ij}/r_{\varepsilon})^2)$, and the QTC consensus matrix
      $C_\text{LGG/GBM}$  calculated assuming a different number of
      clusters. Both weighted and
      unweighted similarity matrices were $\tanh$-normalized.}
    \label{fig:lgggbm}
\end{figure}

The TCGA somatic copy number alteration (SCNA) data in low-grade
glioma (LGG) and glioblastoma (GBM) patient samples were downloaded
from the GDC Data Portal under the name ``LGG/GBM somatic copy number
alterations.''  To link these data to chromatin contact information,
we followed the analysis described in \cite{Fudenberg:2011dv}.  We
partitioned the genome into 1Mb bins and defined $N$ to be a null
square matrix of dimension equal to the total number of bins. For each
amplified or deleted genomic segment starting at the $i$-th bin and
ending at the $j$-th bin, we then incremented the $(i,j)$-th entry of
$N$ by 1. The main idea behind this analysis is that genomic amplification and
deletion events are mediated by the physical co-location of the
segment junctions. The raw count matrix $N$ was thus to be compared with the HiC
chromatin contact matrix. 
In cancer samples, however, 
an entire arm of a chromosome or even a whole chromosome can be duplicated or
deleted, potentially leading to fictitious long-range off-diagonal
elements in $N$.
Therefore, we weighted the counts $N_{ij}$ by
$w_{ij}=\exp[-(r_{ij}/r_\varepsilon)^2]$ where $r_{ij}$ is the genomic
distance between the bins and $r_{\varepsilon} = 10$Mb.  Using this
weighted matrix as an adjacency matrix,  we performed QTC with
$s=5(E_1 - E_0)$, assuming the number of clusters to be $q = 2, 3, 4,
5$, and computed the respective consensus matrices $C(q)$.  Finally, we
took the arithmetic mean $\langle C \rangle = \sum_{q=2}^5 C(q)/4$.

The HiC data in normal human astrocytes of the cerebellum (glial
cells) were downloaded from ENCODE under the name ``ENCSR011GNI''
\cite{Dunham:2012short}.  We extracted the 3D interaction maps on
chromosome 2 at 1Mb resolution.  The distribution of HiC contact
matrix entries was highly heavy-tailed. In order to compare
$C_\text{HiC}$ with $\langle C_{ij} \rangle \in[0,1]$, we transformed
$C_\text{HiC}$ using $\tanh(C_\text{HiC}/\bar C_\text{HiC}) \in
[0,1)$, where $\bar C_\text{HiC}$ was the mean of all $C_\text{HiC}$
entries. Next, we computed the Pearson correlation coefficients
between the transformed $C_\text{HiC}$ and averaged $\langle C(q)_{ij}
\rangle$.

%

\section{Comparison with other methods}
In this section, we first discuss spectral embedding and then derive
three additional (dis)similarity measures using quantum
mechanics. These measures can be combined with spectral clustering as
well as other (dis)similarity-based learning algorithms.

\subsection{Spectral embedding \label{sec:spectral_embedding}}
The state-of-the art spectral clustering can be decomposed into three
major steps: (1) assemble an affinity matrix $A$ based on some
similarity measure of sample points, (2) compute the symmetric
normalized graph Laplacian $H$, and (3) map each sample point indexed
by $i=0,1,\ldots,m-1$ to a Euclidean feature space using the
corresponding elements of eigenvectors of the graph Laplacian; this
mapping is called the spectral embedding.  The first two steps are
essentially the same as those of QTC; the key difference lies in the
final usage of ``spectral properties'' of the data set.  A single
iteration of QTC succinctly represents the data on $S^1$, which we
have shown is sufficient to separate distinct clusters.

By contrast, spectral embedding maps data samples to $\mathbb R^q$,
where $q$ is the number of putative clusters, or the number of low
energy states if all putative clusters are clearly separable; then,
the algorithm performs clustering, e.g.~using $k$-means in the feature
space $\mathbb R^q$. The feature vector $\mathbf v_i$ associated with
the $i$-th sample has elements
$$
	(\mathbf v_i)_n =  \psi_n(i) = \langle i | \psi_n \rangle,\quad n=0,1,\ldots,q-1,
$$
where the $\psi_n$'s are the first $q$ lowest-eigenvalue eigenvectors of $H$.
The $L^2$ Euclidean distance between nodes $(i,j)$ is then
\begin{align}
	{\cal D}_{ij} = & \sqrt{ \| \mathbf v_i - \mathbf v_j\|^2 } \notag\\
	=& \sqrt{\sum_{n=0}^{q-1} |\psi_n(i) - \psi_n(j)|^2} \notag\\
	=& \sqrt{\sum_{n=0}^{q-1} (\langle i | -\langle j |)|n\rangle
          \langle n | (|i \rangle - |j \rangle)}\, .
\end{align}
Note that if we actually used all eigenvectors of $H$, then ${\cal
  D}_{ij} = \sqrt{2(1-\delta_{ij})}$, i.e.~each point is equally far
away from any other node. Thus, the useful clustering information originates
from the projection to low energy states,
\begin{align}
	{\cal D}_{ij}
	=& \sqrt{ (\langle i | -\langle j |) \mathcal P_{n<q} (|i \rangle - |j \rangle)} \\
	\equiv & \sqrt{\chi_{ii}+\chi_{jj} - \chi_{ij} - \chi_{ji}},
\end{align}
where $\chi_{ij} = \langle i | \mathcal P_{n<q} | j \rangle \equiv
\sum_{n<q}\psi_n(i) \psi^*_n(j)$.

In real data, the number of nodes as well as the distribution of node
density could vary from one cluster to another. If a network is
embedded in $\mathbb R^d$, then high density regions contain hub
nodes, provided the adjacency $A_{ij}$ is measured with a non-negative
function that decreases with increasing distance $r_{ij}$,
e.g.~Gaussian function $A_{ij} = \exp(-r_{ij}^2/r_\varepsilon^2)$. For
networks not embedded in $\mathbb R^d$, the ``density'' distribution
should be interpreted as the degree distribution. We next illustrate
how the spectral embedding distance $\mathcal{D}_{ij}$ responds to
outliers in the presence of density variations using
the simple two-cluster model.

\begin{figure}[htb]
    \centering
    \includegraphics[width=1.0\columnwidth]{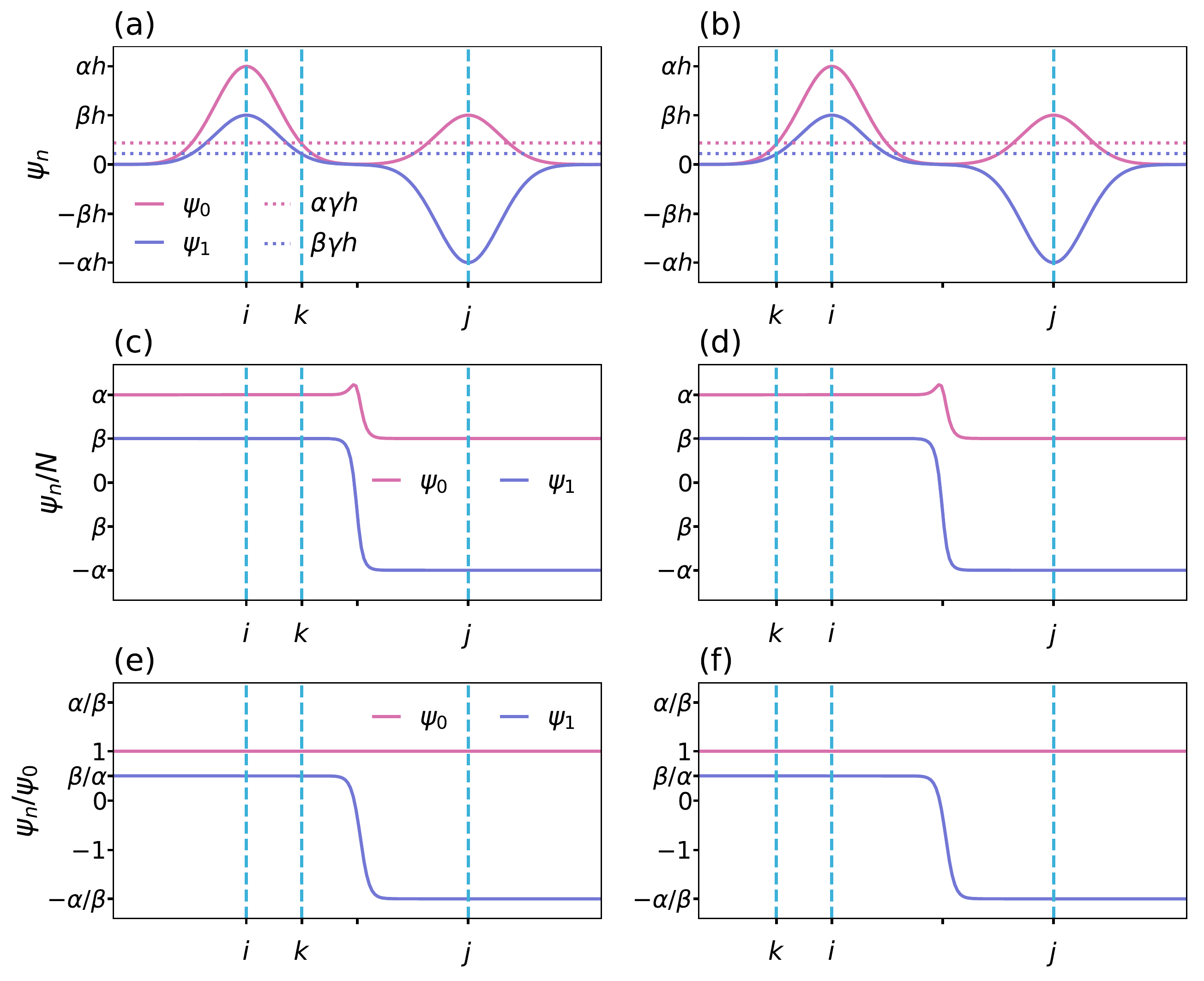}
    \caption{(a,b) Schematic illustrations of the ground state and the
      first excited state involving two clusters; $i,j$, and $k$ are
      node indices. Node $k$ is an outlier (a) lying between the two
      clusters or (b) far from both clusters. (c,d) The normalized
      ground state and first excited state eigenfunctions using
      Approach 1. (e,f) The modified ground state and first excited
      state eigenfunctions using Approach 2.}
    \label{fig:spectral_embedding}
\end{figure}

Using the same notation as in the main text, the ground state and
first excited state, shown in Fig.~\ref{fig:spectral_embedding}(a,b),
are $\psi_{0}=\alpha\phi_{0}+\beta\phi_{1}$ and
$\psi_{1}=\beta\phi_{0}-\alpha\phi_{1}$, where $\alpha,\beta > 0$, and
$\alpha^2 + \beta^2 = 1$.  If we assume $\phi_0$ and $\phi_1$ are
orthonormal,
i.e.~$\langle\phi_{\mu}|\phi_{\nu}\rangle=\delta_{\mu\nu}$ for
$\mu,\nu=0,$ and $1$, then
$\langle\psi_{n}|\psi_{n'}\rangle=\delta_{nn'}$ for $n,n'=0,$ and $1$.
To simplify calculations, we further assume that $\phi_{0}$ and
$\phi_{1}$ have identical shapes with the maximum value $h$ located at
node $i$ and $j$, respectively; i.e.~$\phi_{0}(i)=h=\phi_{1}(j)$.
Then, $\psi_{0}(i)=\alpha h = -\psi_{1}(j)$ and $\psi_{1}(i)=\beta h =
\psi_{0}(j)$.  Let $\gamma\in(0,1]$ such that $\phi_0(k) = \gamma
\phi_0(i) = \gamma h$.  Then, $\psi_0(k) = \gamma \alpha h$, and
$\psi_1(k) = \gamma \beta h$
(Fig.~\ref{fig:spectral_embedding}(a,b)). Recall that $\psi_0 (i) =
\sqrt{{\rm deg}(i)}$ for a normalized symmetric Laplacian; hence, the
differences in $\psi_0$ across nodes can be viewed as capturing the
density variations in a network.

Simple calculations show that
\[
\chi_{ii}=\chi_{jj}=h^{2}\left(\alpha^{2}+\beta^{2}\right) = h^2
\]
\[
\chi_{ij}=\chi_{ji}=h^{2}\left(\alpha\beta-\beta\alpha\right)=0
\]
\[
\chi_{kk}=\left(\gamma h\right)^{2}\left(\alpha^{2}+\beta^{2}\right) = \gamma^2 h^2
\]
\[
\chi_{ik}=\chi_{ki}=\gamma h^{2}\left(\alpha^{2}+\beta^{2}\right) = \gamma h^2
\]
and
\[
\chi_{jk}=\chi_{kj}=\gamma h^{2}\left(\alpha\beta-\beta\alpha\right)=0.
\]
Hence, we find 
\begin{align}
{\cal D}_{ij} & =\sqrt {2}\, h\\
{\cal D}_{ik} & =(1-\gamma) h \\
{\cal D}_{jk} & = \sqrt{1+\gamma^{2}} \,h
\end{align}
with
\[
{\cal D}_{ij}\ge{\cal D}_{jk}>{\cal D}_{ik} \text{ for } \gamma\in(0,1].
\]
In the limit $k$ becomes an outlier of the left cluster $\phi_1$,
$\gamma \downarrow 0$ and ${\cal D}_{ik} \approx {\cal D}_{jk}$.
Furthermore, although the inequalities ${\cal D}_{ij}>{\cal D}_{ik}$
and ${\cal D}_{jk}>{\cal D}_{ik}$ facilitate the task of grouping
similar points, the inequality ${\cal D}_{jk}\le{\cal D}_{ij}$ could
potentially undermine the clustering accuracy.  Notice that node $k$
can be either close or far from the right cluster
(Fig.~\ref{fig:spectral_embedding}(a,b), respectively), but yield the
same ${\cal D}_{jk}$, as long as $\phi_{\mu}(k)=\gamma\phi_{\mu}(i)$. In
other words, an outlier from the left cluster could be closer to the
right cluster in spectral distance, even when the outlier has a
negligible connection to the right cluster
(Fig.~\ref{fig:spectral_embedding}(b)).  By sharp contrast, in QTC,
the phase at a node lying between two clusters interpolates
monotonically between the phases of the two clusters
(Fig. \ref{fig:outlier_o}).

This undesirable behavior of spectral clustering may be avoided by 
renormalizing the eigenvectors.  
Two common approaches are 
(Fig.~\ref{fig:spectral_embedding}(c,d) and (e,f), respectively):
\begin{enumerate}
	\item[] \hspace{-4mm}\underline{\bf Approach 1}
	\item Compute $N(i) \equiv (\sum_{n=0}^{q-1} |\psi_n(i)|^2)^{\frac12}$.
	\item Divide each $\psi_n(i)$ by $N(i)$, i.e.~$\psi_n \rightarrow \psi_n /N$.
\end{enumerate}

\begin{enumerate}
	\item[] \hspace{-4mm}\underline{\bf Approach 2}
	\item Divide each $\psi_n(i)$ by $\psi_0 (i)$, i.e.~$\psi_n
          \rightarrow \psi_n /\psi_0$.
\end{enumerate}
Similar to the phase plateaus in QTC, $\psi_n/N$ and $\psi_n/\psi_0$
are essentially flat within a cluster
(Fig.~\ref{fig:spectral_embedding}(c,d) and (e,f), respectively).

In the first approach (Fig.~\ref{fig:spectral_embedding}(c,d)), the spectral embedding distances become
\begin{align}
{\cal D}_{ij}^{(1)} & = \sqrt{(\alpha - \beta)^2 + (\alpha + \beta)^2 } = \sqrt 2\\
{\cal D}_{ik}^{(1)} & = 0 \\
{\cal D}_{jk}^{(1)} & = \sqrt{(\alpha - \beta)^2 + (\alpha + \beta)^2 } = \sqrt 2.
\end{align}
In the second approach (Fig.~\ref{fig:spectral_embedding}(e,f)), the spectral embedding distances become
\begin{align}
{\cal D}_{ij}^{(2)} & = \sqrt{(\beta/\alpha + \alpha/\beta)^2 } = 1/\alpha\beta \\
{\cal D}_{ik}^{(2)} & = 0 \\
{\cal D}_{jk}^{(2)} & = \sqrt{(\beta/\alpha + \alpha/\beta)^2 } = 1/\alpha\beta.
\end{align}
In both cases, we have ${\cal D}_{jk}^{(1,2)} = {\cal
  D}_{ij}^{(1,2)}$; thus, the outlier node $k$ is much more likely to be
clustered with the left cluster.
(Scikit-Learn, a very popular machine learning
software package in Python, implements the second approach incorrectly as
$\psi_n \rightarrow
\psi_n \times \psi_0$ and sometimes yields  counter-intuitive
clustering results.  In this paper, we use our own implementation of
Approach 1.)

Finally, we note that spectral embedding has an intrinsic weakness
stemming from ignoring potentially useful information from high-energy
states. More precisely, recall that spectral embedding assumes that
the most relevant information for clustering is encoded in the first
$q$ low-energy eigenstates of $H$. However, this assumption could be
invalid in some cases, e.g.~our synthetic data sets in Fig.~1, and
time series data in Fig.~2, where the information needed to separate
some small clusters are stored in higher energy modes. In such a case,
spectral clustering may not have the required information to separate
the small clusters, but instead chop the large clusters into fragments
at their weak edges in low density regions. By contrast, QTC does not
require a manual cut-off in the spectrum and incorporates all
eigenstates by naturally weighing the contribution from each
eigenfunction $\psi_n$ by $|s+{\rm i}E_n|^{-1}$. This difference may
explain why QTC is more robust than spectral embedding when there
exists a hierarchy of cluster sizes.

\begin{figure}[t]
    \centering
    \includegraphics[width=1.0\columnwidth]{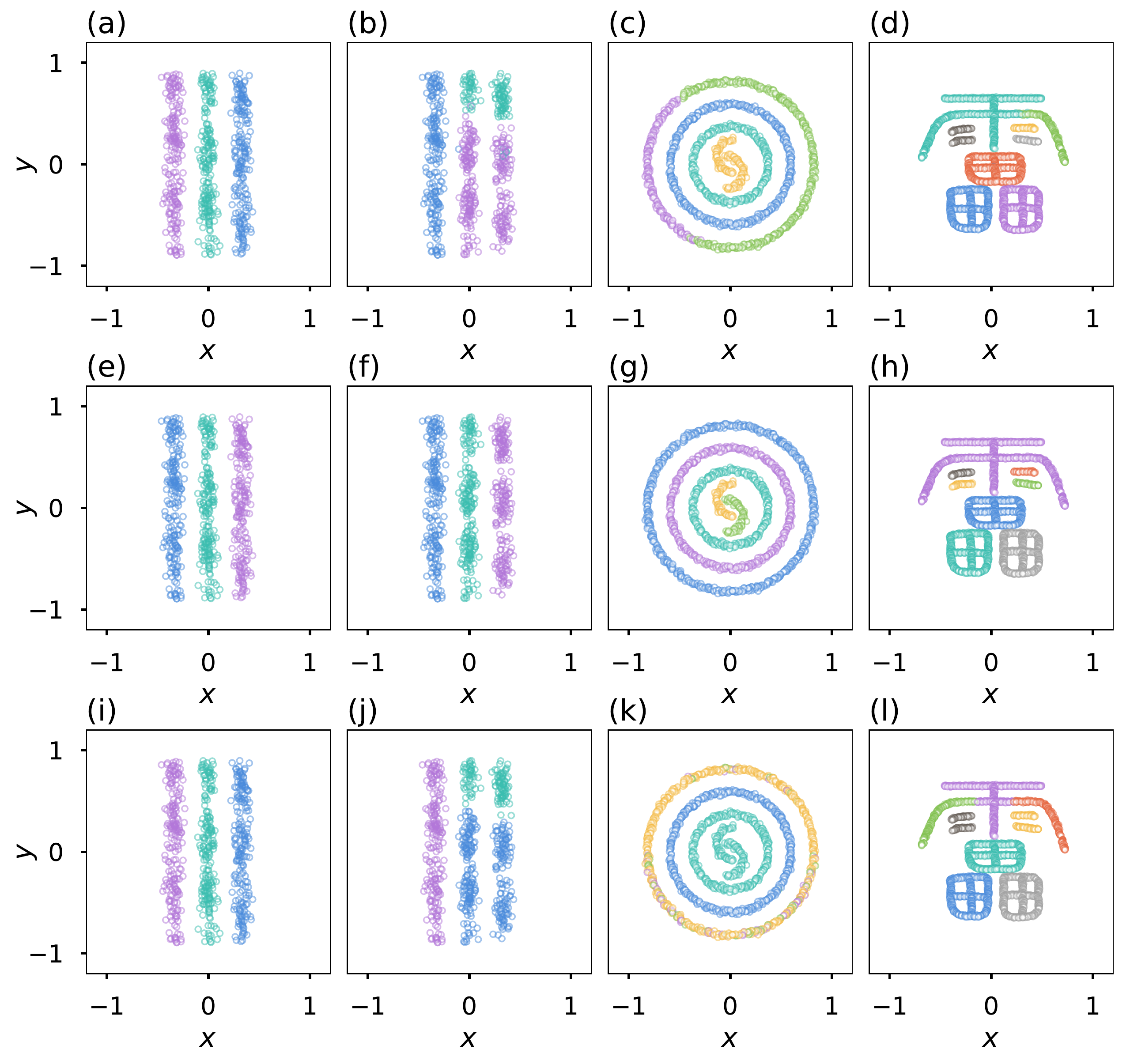}
    \caption{Synthetic data distributions plotted in Fig.~1. Spectral
      clustering was performed using as a similarity measure  (a-d) the 
      time-averaged squared transition amplitude, (e-h) the consensus matrices
      $C$ produced by QTC, and (i-k) the
      similarity $S$ of Laplace-transformed wave functions.}
    \label{fig:syn_data}
\end{figure}

\subsection{Time-averaged transition amplitude}

The time-dependent transition amplitude $G_{ij}(t)$ from node $j$
to $i$ is complex-valued and oscillatory in time, i.e.
\begin{eqnarray*}
G_{ij}(t) & = & \langle i|{\rm e}^{-{\rm i}Ht}|j\rangle\\
 & = & \sum_{m,n}\langle i|\psi_m\rangle\langle \psi_m|{\rm e}^{-{\rm i}Ht}|\psi_n\rangle\langle \psi_n|j\rangle\\
 & = & \sum_{n}\psi_{n}(i)\psi_{n}^{*}(j){\rm e}^{-{\rm i}E_{n}t}.
\end{eqnarray*}
To obtain a real-valued matrix, we take the squared amplitude,
\begin{eqnarray*}
|G_{ij}(t)|^{2} & = & G_{ji}(-t)G_{ij}(t)\\
 & = & \sum_{m,n}\psi_{m}(j)\psi_{m}^{*}(i)\psi_{n}(i)\psi_{n}^{*}(j){\rm e}^{{\rm i}(E_{m}-E_{n})t}\\
 & = & \sum_{m,n}\rho_{mn}(i)\rho_{nm}(j){\rm e}^{{\rm i}(E_{m}-E_{n})t}
\end{eqnarray*}
where $\rho_{mn}(i)=\langle \psi_m|i\rangle\langle i|\psi_n\rangle$. The oscillation
in time can be averaged as
\begin{eqnarray*}
P_{ij} & = & \lim_{T\uparrow\infty}\frac{1}{T}\int_{0}^{T}dt\:|G_{ij}(t)|^{2}\\
 & = & \sum_{m,n}\rho_{mn}(i)\rho_{nm}(j)\left[\lim_{T\uparrow\infty}\frac{1}{T}\int_{0}^{T}dt\:{\rm e}^{{\rm i}(E_{m}-E_{n})t}\right]\\
 & = & \sum_{m,n}\delta_{E_{m},E_{n}}\rho_{mn}(i)\rho_{nm}(j).
\end{eqnarray*}
If there is no degeneracy in the spectrum of $H$, then the time-averaged
squared transition amplitude simplifies to
\begin{eqnarray*}
P_{ij} & = & \sum_{n}\rho_{nn}(i)\rho_{nn}(j)=\sum_{n}|\psi_{n}(i)|^{2}|\psi_{n}(j)|^{2},
\end{eqnarray*}
which is a symmetric, non-negative matrix that can be used as a
similarity measure. 

The performance of $P_{ij}$ as a spectral clustering affinity matrix
was tested in four synthetic data sets (Fig.~\ref{fig:syn_data}(a-d))
as well as the stock price time series data
(Fig.~\ref{fig:stock_data}(b)). The performance was similar to
spectral clustering using Gaussian affinity.


\subsection{Density information of Laplace-transformed wave functions}

\begin{figure}[t]
    \centering
    \includegraphics[width=1.0\columnwidth]{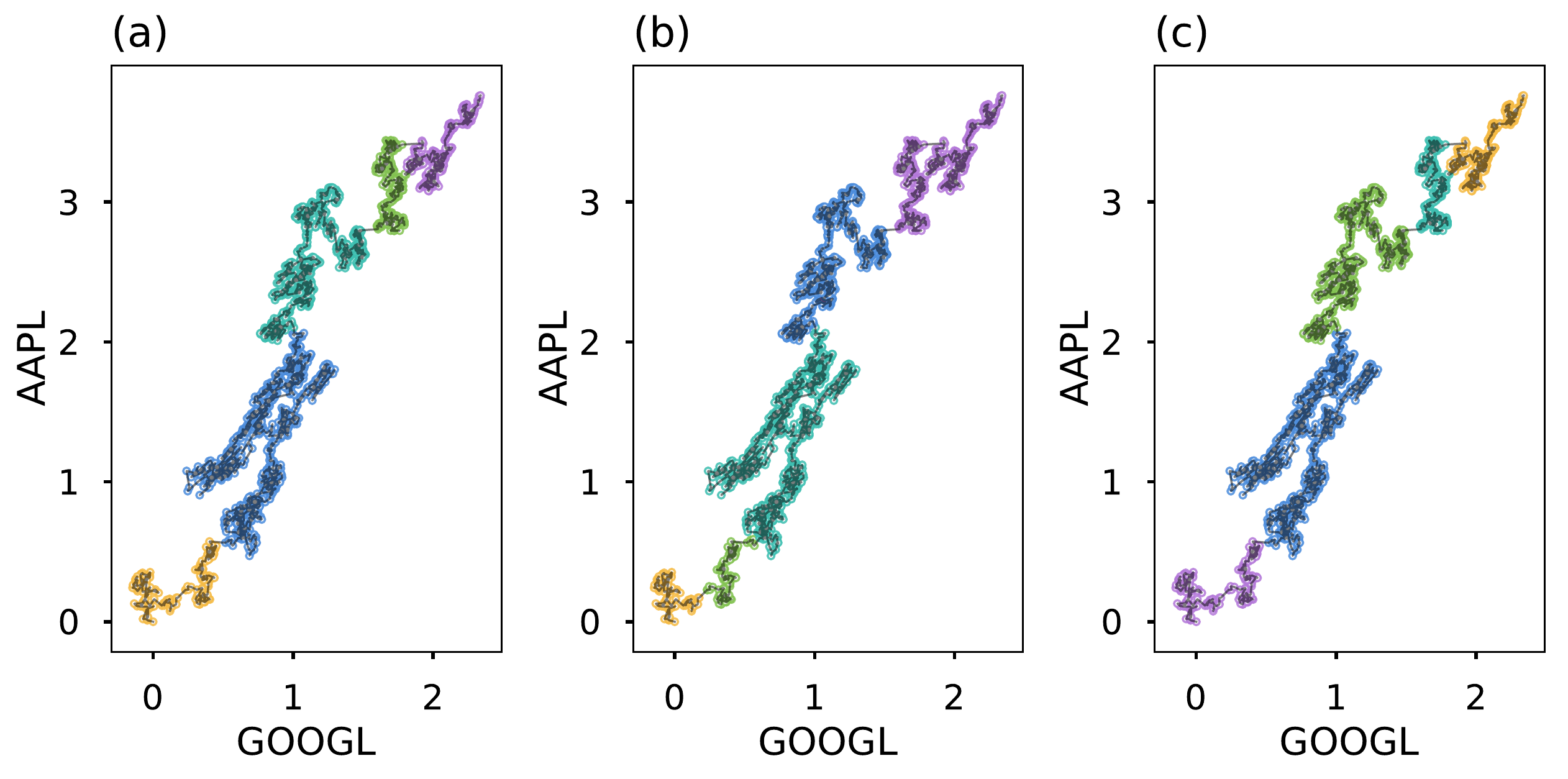}
    \caption{Time series data of the log-prices of AAPL and GOOGL
      stocks from January 1, 2005 to November 7, 2017. Spectral
      clustering was performed using as a similarity measure
      (a) the QTC consensus matrix $C$, (b) the time-averaged squared transition
      amplitude $P$, and (c) the similarity $S$ of Laplace-transformed wave functions.}
    \label{fig:stock_data}
\end{figure}

As in QTC, given a time-independent Hamiltonian, we take the Laplace transform of two
wave functions evolved from  the states initialized at nodes
$i$ and $j$. Then, we take their inner product
\begin{align*}
\langle\tilde{\psi}_{i}(s)|\tilde{\psi}_{j}(s)\rangle & =\langle i|(s-{\rm i}H)^{-1}(s+{\rm i}H)^{-1}|j\rangle\\
 & =\sum_{n}\frac{\psi_{n}(i)\psi_{n}^{*}(j)}{s^{2}+E_{n}^{2}}.
\end{align*}
Next, we define a similarity measure using the inner product
\begin{equation}
  S_{ij} = \left[\frac{\left|\langle\tilde{\psi}_{i}(s)|\tilde{\psi}_{j}(s)\rangle\right|^{2}}{\left|\langle\tilde{\psi}_{i}(s)|\tilde{\psi}_{i}(s)\rangle\right|\left|\langle\tilde{\psi}_{j}(s)|\tilde{\psi}_{j}(s)\rangle\right|}\right]^{\frac{1}{2}},
\end{equation}
which is symmetric and non-negative.  The performance of $S_{ij}$ as
a spectral clustering affinity matrix was also tested on four synthetic
data sets (Fig.~\ref{fig:syn_data}(i-l)) and the stock price time series
data (Fig.~\ref{fig:stock_data}(c)). The performance was similar to that of spectral clustering using Gaussian affinity (Fig.~\ref{fig:syn_data}(i,j,l) and Fig.~\ref{fig:stock_data}(c)), but gave sup-optimal
clustering results on the annulus data set
(Fig.~\ref{fig:syn_data}(k)).

\subsection{Jensen-Shannon divergence of density operators}

The time evolution of the density operator
$\rho(j)=|j\rangle\langle j|$ describing a pure state localized at node
$j$  at time $t=0$ is
\begin{eqnarray*}
\rho(j;t) & = & {\rm e}^{-{\rm i}Ht}|j\rangle\langle j|{\rm e}^{{\rm i}Ht}\\
 & = & \sum_{m,n}{\rm e}^{-{\rm i}Ht}|\psi_m\rangle\left\{ \langle \psi_m|j\rangle\langle j|\psi_n\rangle\right\} \langle \psi_n|{\rm e}^{{\rm i}Ht}\\
 & = & \sum_{m,n}{\rm e}^{-{\rm
     i}(E_{m}-E_{n})t}\rho_{mn}(j)|\psi_m\rangle\langle \psi_n|\, ,
\end{eqnarray*}
where $\rho_{mn}(i)=\langle \psi_m|i\rangle\langle i|\psi_n\rangle$.
If we again take the time average, then
\begin{eqnarray*}
\bar{\rho}(j) & = & \lim_{T\uparrow\infty}\int_{0}^{T}dt\:\rho(j;t)\\
 & = & \sum_{m,n}\delta_{E_{m},E_{n}}\rho_{mn}(j)|m\rangle\langle n|;
\end{eqnarray*}
and, in the absence of energy degeneracy, the time-averaged density operator
initiated at node $j$ simplifies to
\begin{eqnarray*}
\bar{\rho}(j) & = & \sum_{n}\rho_{nn}(j)|\psi_n\rangle\langle
\psi_n|=\sum_{n}|\psi_{n}(j)|^{2}|\psi_n\rangle\langle \psi_n|.
\end{eqnarray*}
For two time-averaged density operators corresponding to pure states
initialized at node $i$ and $j$, respectively,
we may measure the information-theoretic divergence between $\bar{\rho}(i)$ and
$\bar{\rho}(j)$ using  the Jensen-Shannon divergence (JSD),
\begin{eqnarray*}
{\cal D}_{{\rm JS}}[\bar{\rho}(i),\bar{\rho}(j)] & = & {\cal S}\left[\frac{\bar{\rho}(i)+\bar{\rho}(j)}{2}\right]-\frac{1}{2}{\cal S}[\bar{\rho}(i)]-\frac{1}{2}{\cal S}[\bar{\rho}(j)]
\end{eqnarray*}
where ${\cal S}[\rho]=-{\rm Tr}(\rho\log\rho)$ is the von Neumann entropy
of $\rho$. 

\begin{widetext}
Using the eigenfunctions of $H$, 
	\begin{eqnarray*}
{\cal D}_{{\rm JS}}[\bar{\rho}(i),\bar{\rho}(j)] & = & \sum_{n}\left\{ -\frac{|\psi_{n}(i)|^{2}+|\psi_{n}(j)|^{2}}{2}\log\frac{|\psi_{n}(i)|^{2}+|\psi_{n}(j)|^{2}}{2}+\frac{1}{2}|\psi_{n}(i)|^{2}\log|\psi_{n}(i)|^{2}+\frac{1}{2}|\psi_{n}(j)|^{2}\log|\psi_{n}(j)|^{2}\right\} 
\end{eqnarray*}
\end{widetext}
which is a non-linear function of $|\psi_{n}|^{2}$. The time-complexity
for tabulating all elements in pairwise JSD matrix scales as ${\cal
  O}(m^{3})$, where $m$ is the total number of nodes, and the
computation is very slow compared with the proposed QTC 
method. Using small synthetic data sets, we nevertheless implemented the JSD method
and passed the JSD matrix to hierarchical clustering as a dissimilarity
measure. The JSD measure did not show  a significant performance improvement
compared with the simple Euclidean distance.

\bibliographystyle{apsrev4-1}

\end{document}